\documentclass[11pt,a4paper]{article}
\pdfoutput=1
\usepackage{jheppub}
\usepackage{amsfonts, amssymb, amsmath}
\usepackage{mathrsfs}
\usepackage{graphicx}
\usepackage[utf8]{inputenc}
\usepackage[russian,english]{babel}
\usepackage{slashed}

\usepackage{color}
\definecolor{dark-gray}{gray}{0.20}
\definecolor{gray}{gray}{0.30}
\definecolor{light-gray}{gray}{0.80}
\definecolor{dark-red}{rgb}{0.7,0,0}
\definecolor{dark-green}{rgb}{0.1,0.4,0}
\definecolor{dark-blue}{rgb}{0.3,0.3,0.7}
\definecolor{light-blue}{rgb}{0.8,0.8,1}
\definecolor{blue}{rgb}{0,0,1}
\definecolor{red}{rgb}{1,0,0}
\definecolor{green}{rgb}{0,1,0}

\usepackage{hyperref}
\hypersetup{
	colorlinks=true,
	linkcolor=dark-blue,
	citecolor=dark-red,
	urlcolor=dark-blue,
	linktoc=page
}

\def\cB{{\cal B}}

\def\cF{{\cal F}}

\def\cK{{\cal K}}
\def\cL{{\cal L}}

\def\cN{{\cal N}}

\def\cR{{\cal R}}

\def\cV{{\cal V}}

\def\cG{{\cal G}}
\def\cC{{\cal C}}
\def\cX{{\cal X}}

\def\SO{{\rm SO}}
\def\U{{\rm U}}
\def\SU{{\rm SU}}

\def\i{{\rm i}}

\newcommand{\be}{\begin{equation}}
\newcommand{\ee}{\end{equation}}
\newcommand{\bea}{\begin{eqnarray}}
\newcommand{\eea}{\end{eqnarray}}

\title{Maximally symmetric nuts in\\ 4d $\cN=2$ higher derivative supergravity}

\author{Kiril Hristov}

\affiliation{Faculty of Physics, Sofia University, J. Bourchier Blvd. 5, 1164 Sofia, Bulgaria}
\affiliation{INRNE, Bulgarian Academy of Sciences, Tsarigradsko Chaussee 72, 1784 Sofia, Bulgaria}

\emailAdd{khristov@phys.uni-sofia.bg}

\abstract{
\noindent We initiate a systematic study of supersymmetric backgrounds in 4d $\cN=2$ Euclidean supergravity in the presence of infinite towers of higher derivative corrections. Adopting a Gibbons-Hawking view towards the evaluation of the action in terms of nuts and bolts, we consider the two maximally symmetric vacua $\mathbb{R}^4$ and $\mathbb{H}^4$ (Euclidean AdS$_4$) and their unique supersymmetric deformations with (anti-) self-dual Maxwell tensors corresponding to a single nut at the center. These are the Omega background of Nekrasov-Okounkov, $\Omega\, \mathbb{R}^4$, and its generalization with a cosmological constant of Martelli-Passias-Sparks, denoted $\Omega\, \mathbb{H}^4$ (also known as the gravity dual of the $\U(1) \times \U(1)$ squashed sphere). We write down the BPS configurations in the superconformal formalism in the presence of vector multiplets and derive the corresponding off- and on-shell actions. Our results provide a rigorous proof for important parts of the conjecture in \cite{Hristov:2021qsw} and its holographic corollary in \cite{Hristov:2022lcw}, which we discuss in detail along with extensions such as the addition of hypermultiplets and the presence of conical defects.
}
\date{\today}
\begin{document}
\maketitle


\section{Introduction and main results}
\label{sec:intro}
Gravitational instantons, i.e.\ classical solutions of Euclidean gravity and supergravity, have been known for a long time to play a crucial role in the gravitational path integral and the thermodynamic understanding of black holes, see \cite{Gibbons:1976ue}. Under the mild assumption for the existence of at least one continuous symmetry, Gibbons and Hawking further showed  the appeal of classifying such solutions in terms of the fixed point set of this isometry, \cite{Gibbons:1979xm}, with fixed points referred to as ``nuts'' and fixed two-submanifolds as ``bolts''. They showed that the description in terms of nuts and bolts not only allows for the simple evaluation of purely topological invariants such as the Euler characteristic and the signature, but also facilitates the calculation of physical observables such as the on-shell action. Negatively curved solutions were originally not investigated as they were considered unphysical at the time, but the development of the AdS/CFT correspondence prompted renewed interest in the subject, see \cite{Chamblin:1998pz,Martelli:2012sz,Farquet:2014kma,BenettiGenolini:2016tsn,Toldo:2017qsh}. These investigations ultimately allowed the direct evaluation of the on-shell action of all supersymmetry-preserving asymptotically locally AdS backgrounds in terms of their nuts and bolts, as shown in \cite{BenettiGenolini:2019jdz} in minimal (two derivative) gauged 4d $\cN=2$ supergravity.~\footnote{The preservation of supersymmetry guarantees a continuous isometry for all backgrounds, automatically fulfilling the original assumption of Gibbons and Hawking.}

In an {\it a priori} far removed (both in time and in scope) development, observables in conformal and supersymmetric (non-gravitational) field theories have been shown to factorize in terms of basic building blocks in many cases, see \cite{Belavin:1984vu,Nekrasov:2002qd,Nekrasov:2003vi,Beem:2012mb,Festuccia:2018rew} and references thereof, as often shown via supersymmetric localization \cite{Pestun:2007rz,Pestun:2016zxk}. Based on the AdS/CFT correspondence, the 3d  SCFT factorization in terms of holomorphic blocks, \cite{Beem:2012mb}, inspired the construction of {\it gravitational blocks} in (two derivative) 4d $\cN=2$ supergravity coupled to matter, \cite{Hosseini:2019iad}, applied to supersymmetric black hole backgrounds. At this stage the relation to the original classification of Gibbons-Hawking becomes rather obvious, given the fact that the gravitational blocks can be seen to appear precisely at fixed points of the spacetime isometry following from supersymmetry. The additional process of {\it gluing} of these blocks, which (in field theory language) is the identification of the Coulomb branch parameters on the different fixed points of the given background, can be seen as an effect from the generalization of the Gibbons-Hawking construction to theories with extra matter.

In the present work we try to bridge the remaining gap between these two pictures by considering the simplest backgrounds exhibiting a single nut, i.e.\ we deform in a supersymmetry-preserving way the maximally symmetric Euclidean backgrounds $\mathbb{R}^4$ and $\mathbb{H}^4$ (or Euclidean AdS$_4$). As we show explicitly, keeping the same background metric still allows the addition of a non-trivial self-dual or anti-self-dual two-form field and a corresponding set of gauge fields (which do break some of the underlying isometries of the metric but importanltly keep a $\U(1) \times \U(1)$ invariance). In field theory language this is known as $\Omega$-deformation \cite{Nekrasov:2002qd}, and the corresponding background (denoted here as $\Omega\, \mathbb{R}^4$) was first introduced by Nekrasov-Okounkov in \cite{Nekrasov:2003rj}. At the same time we are interested in holographic application and are therefore driven to put further emphasis on the direct generalization of this deformation on $\mathbb{H}^4$, which was first constructed in minimal Euclidean supergravity by Martelli-Passias-Sparks~\footnote{This background can be obtained by Wick rotation of the Plebanski-Demianski metric \cite{Plebanski:1976gy} after imposing the supersymmetry limit discussed in \cite{Alonso-Alberca:2000zeh}.} as the gravity dual of the $\U(1) \times \U(1)$-invariant squashed sphere, \cite{Martelli:2011fu}. For reasons that will become clear in due course,~\footnote{Let us emphasize that the general Nekrasov-Okounkov deformation \cite{Nekrasov:2003rj}, or twist, goes beyond flat space and generalizes the Donaldson-Witten twist on four-manifolds with continuous isometries. This deformation is {\it not} possible on $\mathbb{H}^4$ in supergravity (as opposed to rigid supersymmetry following from a frozen gravity multiplet, see \cite{Klare:2013dka}) due to the need for $\SU(2)$ R-symmetry that is broken by the gauging.}  we refer to this background (which we consider in global spherical coordinates) as $\Omega\, \mathbb{H}^4$.

While keeping in mind the above discussion, we should add that another major reason for the present analysis is the plan of systematically addressing and ultimately proving the conjecture in \cite{Hristov:2021qsw} about the form of the action for all supersymmetric backgrounds in the presence of infinite classes of higher derivative terms. We are able to make some important steps towards this goal, while still leaving a number of unfinished tasks for the future, as commented later. Due to our scope we need to use the off-shell superconformal gravity formalism, see \cite{Lauria:2020rhc} for a review, in Euclidean signature \cite{deWit:2017cle}. On top of the standard gauged supergravity at two derivatives, we allow for the Weyl-squared \cite{Bergshoeff:1980is} (here denoted as $\mathbb{W}$) and T-log \cite{Butter:2013lta} (here denoted as $\mathbb{T}$) infinite towers of higher derivative terms. Technically, our analysis of the supersymmetry variations in the presence of hypermultiplets and vector multiplets is to a large extent analogous to the one in \cite{Hristov:2009uj} for maximally supersymmetric spacetimes. Note that (both Lorentzian and Euclidean) supersymmetric backgrounds in 4d $\cN=2$ conformal supergravity were classified in \cite{Klare:2013dka} specifically at the level of the gravity multiplet, with $\Omega\, \mathbb{R}^4$ written down explicitly based on the results in \cite{Hama:2012bg}.~\footnote{It was later pointed out in \cite{Bobev:2019ylk} (pursuing the 5d holographic dual) that this background preserves all $Q$ and half of the $S$ supercharges.} The explicit supersymmetry analysis of the Martelli-Passias-Sparks background seems to appear for first time here in the superconformal setting. In turn our evaluation of the corresponding background actions, using also holographic renormalization, \cite{Gibbons:1979xm,Chamblin:1998pz,Skenderis:2002wp}, is in line with the four derivative computations in \cite{Bobev:2020egg,Bobev:2021oku,Genolini:2021urf} with some important technical differences we describe along the way, most notably the fact that we do not impose the equations of motion until the very end (supersymmetry does impose many of them automatically), keeping the action off-shell. We then arrive at the on-shell action for $\mathbb{H}^4$ and $\Omega\, \mathbb{H}^4$ by simply extremizing the off-shell action with respect to a remaining set of constant scalars, in analogy to $F$-extremization in the holographically dual field theory, \cite{Jafferis:2010un,Jafferis:2011zi}.~\footnote{Note that we do not consider running scalars as in the background in \cite{Freedman:2013oja}, which is a further generalization of $\mathbb{H}^4$ in Euclidean supergravity, but the off-shell formalism allows us to uncover a very similar process of fixing the values of the scalars. See also \cite{Binder:2021euo,Zan:2021ftf} for related observations.}

We can already present some of our main results in a simplified fashion, leaving most of the technical details for the main sections below. If we focus only on the bosonic fields that remain in the on-shell formalism, we have the metric $g_{\mu \nu}$ and the $n_V+1$ ablelian gauge fields $W^I_\mu$, $I = 0, 1, ..., n_V$, along with the same number of complex scalars $X^I$ (the on-shell scalars can be taken as $z^i = X^i / X^0, i \neq 0$). In order to introduce the two towers of higher derivative invariants, we can look at the composite scalars $A_\mathbb{W}$ and $A_\mathbb{T}$ (defined in the main text) corresponding to the $\mathbb{W}$ and $\mathbb{T}$ invariants, respectively. The supergravity Lagrangian is then uniquely specified by the choice of gauging and the holomorphic prepotential function that takes the generic form of a double expansion,~\footnote{Note that in Euclidean signature one can actually choose two independent prepotential functions, $F^+ (X^I_+; A^+_\mathbb{W}, A^+_\mathbb{T})$ and $F^- (X^I_-; A^-_\mathbb{W}, A^-_\mathbb{T})$, since the chiral and anti-chiral multiplets that make up a vector multiplet do not need to be taken in a symmetric way, see \cite{Bobev:2021oku}. Here we are interested in theories with {\it real} action and do not explore this additional freedom. The two independent chiralities of the composite fields $A_\mathbb{W}$ and $A_\mathbb{T}$ do play a prominent role discussed in the technical sections of this work.}
\be
\label{eq:1}
	F (X^I; A_\mathbb{W}, A_\mathbb{T}) = \sum_{m, n = 0}^\infty F^{(m,n)} (X^I)\, (A_\mathbb{W})^m\, (A_\mathbb{T})^n\ ,  
\ee
with each $F^{(m,n)} (X^I)$ a homogeneous function of degree $2 (1- m-n)$.

The $\Omega\, \mathbb{R}^4$ background is shown to be a fully-BPS configuration of ungauged supergravity, given in spherical Hopf coordinates by
\be
		{\rm d} s^2 = {\rm d} r^2 + r^2 \left( {\rm d} \theta^2 + \cos^2 \theta\, {\rm d} \varphi_1^2 + \sin^2 \theta\, {\rm d} \varphi_2^2 \right) \ ,	
\ee
which is the metric on flat space, together with background gauge fields
\be
	W^I =  \frac{X^I}8\,  (\epsilon_1 - \epsilon_2)\,  (r^2  \cos^2 \theta\, {\rm d} \varphi_1 - r^2 \sin^2 \theta\,  {\rm d} \varphi_2)\ ,
\ee
for arbitrary constant scalars $X^I$, giving rise to purely anti-self-dual field strengths, $F^{I +}_{\mu\nu} = 0$ (there is an analogous configuration with only non-vanishing self-dual field strengths after a flip of the relative sign). Note that the field strengths are covariantly constant, and apart from the set of constant scalars $X^I$ we have one additional free deformation parameter, $(\epsilon_1 - \epsilon_2)$, which we have written in the form chosen in the original references, \cite{Hama:2012bg,Klare:2013dka}. This BPS {\it configuration} is shown to automatically be a {\it solution} of all equations of motion, but it turns out that the corresponding on-shell action is {\it not} finite and even upon regularization one cannot extract a meaningful non-vanishing answer. The calculation leading to this result is however very useful to clarify some particularities of asymptotically flat solutions and their place in the conjecture of \cite{Hristov:2021qsw}, which we discuss in section \ref{sec:on}.

The $\Omega\, \mathbb{H}^4$ background is instead a half-BPS configuration in gauged supergravity. In the case of Fayet-Iliopoulos constant gauging parameters (see later for details), labeled by $g_I$, we again find constant scalars $X^I$ with the special linear relation $g_I X^I = L^{-1}$. The metric is then given by
\be
	{\rm d} s^2 = \frac{{\rm d} r^2}{1+\tfrac{r^2}{L^2}} + r^2 \left( {\rm d} \theta^2 + \cos^2 \theta\, {\rm d} \varphi_1^2 + \sin^2 \theta\, {\rm d} \varphi_2^2 \right) \ ,	
\ee
which is the hyperbolic space $\mathbb{H}^4$, with gauge fields
\be
	W^I = -L\, X^I \frac{(b_1 + b_2 \sqrt{1+\tfrac{r^2}{L^2}})\, {\rm d} \varphi_1 +  (b_2 + b_1 \sqrt{1+\tfrac{r^2}{L^2}})\, {\rm d} \varphi_2}{2\, \sqrt{ (b_1 + b_2 \sqrt{1+\tfrac{r^2}{L^2}})^2 \sin^2 \theta +  (b_2 + b_1 \sqrt{1+\tfrac{r^2}{L^2}})^2 \cos^2 \theta}}\ .
\ee
leading again to a purely anti-self-dual field strengths, $F^{I +}_{\mu\nu} = 0$. We again kept the parametrization of the original references, see \cite{Farquet:2014kma}, but it can be seen that the background depends only on a single combination of the parameters $b_1$ and $b_2$. As further elaborated in \cite{Farquet:2014kma}, the background is regular for $b_1/b_2 > 0$, or in the special cases of trivial deformation $b_1 / b_2 = \pm 1$ (in which case $W^I$ is a pure gauge). In the limit $L \rightarrow \infty$, or $g_I = 0, \forall I$, the $\Omega\, \mathbb{H}^4$ configuration becomes equal to the $\Omega\, \mathbb{R}^4$ solution above, provided the simple identification
\be
	\epsilon_{1,2} = \frac{2\, b_{1,2}}{L\, (b_1+b_2)}\ .
\ee
Due to the hyperbolic asymptotics, allowing us to apply holographic renormalization, the action for $\Omega\, \mathbb{H}^4$ is found to be finite. It might be tempting to interpret the $\Omega\, \mathbb{H}^4$ background as the natural way of introducing a UV cut-off to $\Omega\, \mathbb{R}^4$, but this is obscured by various differences between gauged and ungauged supergravity, which we discuss in due course. We find the off-shell action
\be
	S^\text{off-shell}  |_{\Omega\, \mathbb{H}^4}  (X^I, b_1, b_2) = \frac{4 \pi^2}{b_1 b_2} F\left((b_1+b_2) \tfrac{L}{\kappa}\, X^I; (b_1-b_2)^2, (b_1 + b_2)^2\right) \ ,
\ee
after we have restored the Newton constant, $\kappa^2 = 8 \pi\, G_N$, in a meaningful way.~\footnote{Note that in the formalism of superconformal gravity there are a priori no scales, and only the choices of  background and gauge fixing break the additional conformal and superconformal symmetries. The introduction of the Newton constant is therefore a matter of choice and there is no unique description. In the present analysis we stick to the logic outlined in \cite{Hristov:2021qsw} that each higher derivative order in the Lagrangian is suppressed by a respective power of $G_N$.} The Einstein and Maxwell equations are automatically solved by the BPS configuration above, but the scalar equations of motion are only solved upon the extremization of the above action with respect to the parameters $X^I$ under the constraint $g_I X^I = L^{-1}$:
\be
	\frac{\partial S^\text{off-shell}}{\partial X^I} |_{\hat{X}^I}  = 0\ , \qquad S^\text{on-shell} (b_1, b_2) = S^\text{off-shell} (\hat{X}^I, b_1, b_2)\ .
\ee
These results are in precise agreement with the conjecture of \cite{Hristov:2021qsw} and their rigorous derivation in the coming sections can be regarded as a partial proof of it, as we explain in detail. We further elaborate on their holographic significance and their extension in a number of different directions.

The rest of this paper is organized as follows. Our main calculations including all technical details are presented in the next two sections. In section \ref{sec:susy} we introduce the off-shell multiplets and perform the supersymmetry analysis for the backgrounds we consider. The full off-shell configurations for all background fields are {\it independent} of the chosen Lagrangian (but do depend on the gauging choice), such that they can be used for possible newly discovered higher derivative invariants. In section \ref{sec:off} we focus on the Lagrangian that includes the $\mathbb{W}$ and $\mathbb{T}$ towers of higher derivatives as dictated by \eqref{eq:1}, and evaluate the off-shell action for the Omega-deformed backgrounds of interest. We further elaborate on the holographic renormalization procedure that we need for $\Omega\, \mathbb{H}^4$, as well as on the gauge fixing and resulting on-shell actions. Section \ref{sec:on} is devoted to the detailed relation of the present results with the conjecture in \cite{Hristov:2021qsw} and its holographic corollary worked out in \cite{Hristov:2022lcw}. We finally conclude in section \ref{sec:out} with a discussion of open problems and the extension of our results upon the addition of more general hypermultiplet gaugings and the presence of conical singularities on the considered backgrounds.

\section{Supersymmetry analysis}
\label{sec:susy}
As explained above, we are going to be interested in the off-shell supersymmetry conditions for the maximally symmetric spaces $\mathbb{R}^4$ and $\mathbb{H}^4$ and their deformations with (anti-) self-dual tensors that keep the same metric. Due to this setup, our supersymmetry analysis here shares many similarities with \cite{Hristov:2009uj} for maximally supersymmetric Lorentzian configurations, as well as \cite{deWit:2011gk,Hristov:2016vbm,Bobev:2021oku} for supersymmetry in the off-shell formalism in the presence of gauging.

We strictly follow the notation and conventions of \cite{deWit:2017cle} on superconformal gravity in Euclidean signature. Let us note a few particularities about the formalism. The bosonic symmetry algebra consists of general coordinate, local Lorentz, dilatation and special conformal transformations, as well as $\SO(1,1) \times \SU(2)$ R-symmetries, while the fermionic symmetries are the supersymmetry ($Q$) and special supersymmetry ($S$) transformations. The gauge fields associated with the general coordinate transformations (vielbein $e_\mu^a$), dilatations ($b_\mu$), R-symmetry ($A_\mu, \cV_\mu{}^i{}_j$) and $Q$-supersymmetry ($\psi_\mu^i$), are independent fields, while the remaining gauge fields are composite. All these fields, together with an additional number of auxiliary fields, are part of the so-called Weyl multiplet, the superconformal version of the gravity multiplet. In addition, one needs to include a number of so-called compensating multiplets in order for the theory to be gauge-equivalent to the Poincar\'e supergravity. In the present case we include one additional hypermultiplet and one additional vector multiplet to the ones present in the on-shell theory, which we introduce below. An extra feature in Euclidean signature is that the fermions of different chiralities are {\it independent} of each other (in Lorentzian signature they are related via complex conjugation), which in turn means that each supersymmetry variation can be split into an independent chiral and anti-chiral part denoted by $+$ and $-$, respectively. 

Before discussing each type of multiplet separately, note that from the start we set all background fermionic fields to zero, as usual. In addition, we also look for configurations with vanishing conformal and $SO(1,1)$ connections,
\be
\label{eq:fixK}
	b_\mu = A_\mu = 0\ ,
\ee
since non-trivial vevs for them are not of interest here. See \cite{Klare:2013dka} for the role of these fields in the general supersymmetry classification. Below we consider hypermultiplets, vector multiplets, and the Weyl multiplet in the presence of abelian gaugings, and introduce all relevant bosonic fields in the order they appear in the supersymmetry variations.

\subsection{Hypermultiplets}
We begin looking first at superconformal hypermultiplets, which have been described in \cite{deWit:1999fp,deWit:2001bk}, since they allow for a gauging of the internal isometries that in turn generates a scalar potential in supergravity. A very specific feature of hypermultiplets is that, in order to avoid the introduction of an infinite set of fields, supersymmetry is realized on-shell, differently to all other multiplets we consider. The field content of each hypermultiplet is four real scalars $q^u$, packaged inside the pseudo-real local sections $A_i^\alpha$ ($i= 1, 2$ is an $\SU(2)$ index, $\alpha = 1, ..., 2 (n_\mathrm{H}+1)$ for a total of $n_\mathrm{H} +1$ hypermultiplets) and two fermions of each chirality, $\zeta^\alpha_\pm$. The superconformal hypermultiplet geometry is that of a hyper-K\"ahler cone with a metric that follows from the hyper-K\"ahler potential,
\be
	\chi_\mathrm{H} := \frac12 \varepsilon^{i j} \Omega_{\alpha \beta} A_i{}^\alpha A_j{}^\beta \ ,
\ee
which features in the Lagrangian of the theory (see later). The isometries of the underlying geometry can be gauged by the available gauge fields, denoted by $W^I_\mu$, in the vector multiplets, which then naturally must appear in covariant derivatives of hypermultiplet fields. Supersymmetrization further requires that the complex scalars $X^I_\pm$ (see below for more details) feature in the hyperino variation,
\be
\label{eq:hyperino}
	\delta \zeta^\alpha_\pm = -\i \slashed{D} A_i{}^\alpha \epsilon^i_\mp - X^I_\mp t_I^\alpha{}_\beta A_i{}^\beta \epsilon^i_\pm + A_i{}^\alpha \eta^i_\pm\ ,
\ee
where
\be
	D_\mu A_i{}^\alpha = \partial_\mu A_i{}^\alpha -\frac12 \cV_\mu{}^j {}_i A_j{}^\alpha - \frac12 W^I_\mu t_I^\alpha{}_\beta A_i{}^\beta\ .
\ee
The $t_I$ above denote the gauge group generators,\footnote{Note that we have rescaled the gauge group generators with a factor of 2 in comparison with the convention in \cite{deWit:2017cle}.} and we remind the reader that $\cV_\mu{}^i{}_j$ is the gauge field for the $\SU(2)$ R-symmetry. The $Q$-supersymmetry variation is denoted by the spinor $\epsilon^i_\pm$, while the $S$-supersymmetry variation by the spinor $\eta^i_\pm$ where $i$ is an $\SU(2)$ index and $\pm$ denotes chirality.

At this stage we can formally write down the supersymmetry conditions that follow from the assumption of maximal supersymmetry, i.e.\ that different $Q$-spinors are independent of each other. On the other hand, it is inconsistent to set the sections $A_i{}^\alpha$ to zero, and therefore it is clear that the $S$-spinors must be fixed either to zero or in terms of the $Q$-variations, and they do not correspond to independent variations. This is precisely expected to happen in supergravity (as opposed to rigid supersymmetry on curved spaces) since the $S$-supersymmetries need to be gauge fixed in the Poincar\'e frame. We see that this is automatically achieved here by introducing  at least one hypermultiplet (as required in order to recover Poincar\'e supergravity). We find that
\be
\label{eq:genhyp}
	A_i{}^\alpha = const\ , \quad \cV_\mu{}^j{}_i A_j{}^\alpha = - W^I_\mu t_I^\alpha{}_\beta A_i{}^\beta\ , \quad A_i{}^\alpha \eta^i_\pm =  X^I_\mp t_I^\alpha{}_\beta A_i{}^\beta \epsilon^i_\pm\ .
\ee
However, at this stage these equations are rather abstract since they apply to an arbitrary geometry and gauging, and cannot be further simplified before some commitment to the underlying hypermultiplet sector. We are going to discuss more general hypermultiplet gaugings in section \ref{sec:hyp}, but for the time being we are interested in two special cases where we consider only a single compensating hypermultiplet with a flat hyper-K\"ahler cone.

\subsubsection*{No gauging}
The simplest case is when we have no gauging at all, i.e.\ the limit to ungauged supergravity. In this case we simply have $t_I^\alpha{}_\beta = 0$, and without loss of generality we can choose
	\be
		\chi_H^{-1/2} A_i{}^\alpha = \delta_i{}^\alpha\ ,
\ee
leading in turn to solving \eqref{eq:genhyp} as
\be
\label{eq:ungaugedS}
	\cV_\mu {}^j {}_i = 0\ , \qquad \eta^i_\pm = 0\ . 
\ee
In other words, the $S$-variations are set to zero and the background $\SU(2)$ gauge fields is vanishing such that the original R-symmetry remains intact, as expected in ungauged supergravity.

\subsubsection*{FI gauging}
The other important case we consider here is when we turn on constant gauging parameters, known as Fayet-Iliopoulos (FI) gauging. The FI parameters $g_I$ are used to gauge a $\U(1)$ subgroup of the $\SU(2)$ R-symmetry group. Without loss of generality we can pick a particular direction (see \cite{Hristov:2016vbm} for a more detailed discussion)
	\be
\label{eq:FIgauge}
		t_I^\alpha{}_\beta = \i\, g_I\, \sigma_3{}^\alpha{}_\beta\ , 
\ee
where $\sigma_3$ is the corresponding Pauli matrix, and again fix
\be
\label{eq:gaugefix}
	\chi_H^{-1/2} A_i{}^\alpha = \delta_i{}^\alpha\ ,
\ee
such that \eqref{eq:genhyp} leads to
\be
\label{eq:VtoW}
	 \cV_\mu{}^i{}_j  = -\i\, g_I W^I_\mu \sigma_3{}^i{}_j\ , 
\ee
and
\be
\label{eq:s-gauged}
	\eta^i_\pm  = \i\, g_I X^I_\mp\, \sigma_3{}^i{}_j \epsilon^j_\pm\ .
\ee
From the point of view of the Poincar\'e theory, we have gauged a $\U(1)$ subgroup of the $\SU(2)$ R-symmetry group, and broken the other two generators. The $S$-variation parameters are again fixed, but this time in terms of the $Q$-variations, effectively producing a different supersymmetry variation rules between ungauged and gauged supergravity, again as expected. Note that we can also incorporate the case of ungauged case simply by the FI parameter choice $g_I = 0$, $\forall I$.

\subsection{Vector multiplets}
Now we turn our attention to $n_\mathrm{V}+1$ abelian vector mutliplets with corresponding real vector fields $W_\mu^I$ ($I$ labels each different multiplet), pairs of real scalar fields $X^I_+$ and $X^I_-$, Majorana spinors $\Omega^{I, i}_\pm$, and triplets of pseudo-real auxiliary scalars $Y^{I, ij}$. These degrees of freedom follow from the direct compactification of 5d supergravity along a time-like direction as in \cite{deWit:2017cle}, but we should note that it is equally viable to consider a doubling of the degrees of freedom following from Euclideanization of the Lorentzian 4d multiplets as used in \cite{Freedman:2013oja,Bobev:2020pjk}. From the former point of view we should allow for complex vectors $W^I_\mu$, two pairs of independent complex scalars $X^I_\pm$, and a triplet $Y^{I, ij}$ that does not obey any pseudo-reality condition. Here we follow the conventions of \cite{deWit:2017cle}, but it will later be evident that our analysis does not actually need to commit to either of the two pictures, so our final results allow for both interpretations.

The supersymmetry variation of the gaugini is given by
\be
\label{eq:gaugino}
	\delta \Omega^{I, i}_\pm = - 2 \i\, \slashed{\partial} X^I_\pm \epsilon^i_\mp - \frac12 [ F^I_{\mu\nu} - \frac14 X^I_\mp T_{\mu\nu} ]\, \gamma^{\mu\nu} \epsilon^i_\pm + \varepsilon_{kj} Y^{I, ik} \epsilon^j_\pm + 2 X^I_\pm \eta^i_\pm\ ,
\ee
where the two-form field $T_{\mu \nu}$ is part of the Weyl multiplet (see below). Due to the chirality properties of the spinors, it is convenient to also split the two-form fields according to the their self-duality properties, 
\be
	F^{\pm}_{\mu\nu} := \tfrac12 \left( F_{\mu\nu} \pm \tfrac12 \varepsilon_{\mu\nu\rho\sigma} F^{\rho \sigma} \right)\ ,
\ee
where $\varepsilon_{1 2 3 4} = 1$ is the totally anti-symmetric symbol. We then find the useful identities
\be
\label{eq:usefulid}
	 F^{I, \pm}_{\mu\nu} \gamma^{\mu\nu} \epsilon^i_\pm = 0\ , \qquad  T^{\pm}_{\mu\nu} \gamma^{\mu\nu} \epsilon^i_\pm = 0\ .
\ee
Setting the gaugini variation to zero can again be solved uniquely by the assumption that the spinors $\epsilon$ are completely generic, such that each term in \eqref{eq:gaugino} vanishes independently:
\be
\label{eq:genFtoT}
	X^I_\pm = const\ , \qquad F^{I, \pm}_{\mu\nu} = \frac14 X^I_\pm T^\pm_{\mu\nu}\ ,
\ee
and
\be
	 \varepsilon_{kj} Y^{I, ik} \epsilon^j_\pm = - 2 X^I_\pm \eta^i_\pm\ .
\ee
The last equation can only be simplified further if we now specify to the $S$-variations determined above by the hypermultiplet gaugings.

\subsubsection*{No gauging}
In this case we saw that $\eta^i_\pm = 0$, which leads to the simple condition that
\be
	Y^{I, ij} = 0\ .
\ee

\subsubsection*{FI gauging}
Using \eqref{eq:s-gauged}, we find
\be
	Y^{I, ij} = -2 \i\, X^I_+ (g_J X^J_-) \varepsilon^{j k} \sigma_3{}^i{}_k = -2 \i\, X^I_- (g_J X^J_+) \varepsilon^{j k} \sigma_3{}^i{}_k\ ,
\ee
which relates $X^I_+$ and $X^I_-$ upto some overall freedom. It is a gauge choice (see later) to fix 
\be
\label{eq:fixA}
g_I X^I_+ = g_J X^J_- = \frac1{L}\ ,
\ee
leading to
\be
	X^I_+ = X^I_- = X^I\ , \quad \forall I\ .
\ee
We can therefore write 
\be
\label{eq:YtoX}
	Y^{I, ij} = - \frac{2\, \i}{L} X^I \varepsilon^{j k} \sigma_3{}^i{}_k\ , \qquad  Y^I_{ij} = \frac{2\, \i}{L} X^I \varepsilon_{i k} \sigma_3{}^k{}_j\ ,
\ee
where in the last identity we used the pseudo-reality condition together with the identity $\varepsilon^{i k} \varepsilon_{j k} = \delta^i{}_j$. We also find the identity
\be
\label{eq:simplifY}
	Y^{I}_{ij}\, Y^{J, ij} = \frac{8}{L^2}\, X^I X^J\ ,
\ee
that will be useful in the next section.

\subsection{The Weyl multiplet}
The field content of the Weyl multiplet includes the above mentioned gauge fields for the various superconformal symmetries: the vielbein $e^a_\mu$, the $\SU(2)$ gauge field $\cV_\mu{}^i{}_j$ and its corresponding field strength $R(\cV)_{\mu\nu}{}^i{}_j$, the conformal and $\U(1)$ R-symmetry gauge fields $b_\mu, A_\mu$ (that we already set to zero on the backgrounds of interest), as well as two gravitini $\psi^i_\mu$. Additionally, we have a number of auxiliary fields: the antisymmetric tensor $T_{\mu\nu}$ which decomposes into two independent parts as above, $T^\pm_{\mu\nu}$, a scalar $D$ and two symplectic Majorana fermions $\chi^i$ called dilatini. The weights of each of these fundamental fields under the various symmetries are gathered in various tables in \cite{deWit:2017cle} and references therein, which are worth a careful look for the unaccustomed reader. 

We start with the dilatini variation, given by
\be
	\delta \chi^i_\pm = \frac{\i}{24} \gamma^{\mu\nu} \slashed{\nabla} T^\mp_{\mu\nu} \epsilon^i_\mp + D\, \epsilon^i_\pm + \frac16 R(\cV)^\mp_{\mu\nu}{}^i{}_j \gamma^{\mu\nu} \epsilon^j_\pm + \frac{1}{24} T^\mp_{\mu\nu} \gamma^{\mu\nu} \eta^i_\pm\ ,
\ee
where $D$ is the auxiliary scalar and should not be confused with a derivative. If we again take the approach that different terms vanish separately, we arrive at
\be
\label{eq:dilatinosolve}
	\gamma^{\mu\nu} \slashed{\nabla} T^\mp_{\mu\nu} \epsilon^i_\mp = 0\ , \qquad D = 0\ ,
\ee
as well as
\be
	  R(\cV)^\mp_{\mu\nu}{}^i{}_j \gamma^{\mu\nu} \epsilon^j_\pm = - \tfrac14 T^\mp_{\mu\nu} \gamma^{\mu\nu} \eta^i_\pm\ ,
\ee
which holds automatically both in ungauged and in FI gauged supergravity due to \eqref{eq:ungaugedS} and the combination of  \eqref{eq:VtoW}, \eqref{eq:s-gauged} and \eqref{eq:genFtoT}, respectively. Notice that in the first identify in \eqref{eq:dilatinosolve} we did not strip off the spinorial part, since we re going to solve this in two different ways below.

The gravitini variation instead reads
\be
	\tfrac12 \delta \psi_\mu^i{}_\pm = D_\mu \epsilon^i_\pm - \frac{\i}{8} T^{\mp}_{\mu\nu} \gamma^\nu \epsilon^i_\mp -\frac{\i}{2} \gamma_\mu \eta^i_\mp\ ,
\ee
where the covariant derivative is defined as
\be
\label{eq:covderspinor}
	D_\mu \epsilon^i_\pm := ( \partial_\mu - \tfrac14 \omega_\mu{}^{ab} \gamma_{ab} ) \epsilon^i_\pm + \tfrac12 \cV_\mu{}^i{}_j \epsilon^j_\pm\ .
\ee
There are two different strategies to check that the gravitini variation vanishes: either we find the explicit functional dependence of the Killing spinors $\epsilon_\pm^i$ on a given background and check directly that
\be
\label{eq:gravvar}
	D_\mu \epsilon^i_\pm = \frac{\i}{8} T^{\mp}_{\mu\nu} \gamma^\nu \epsilon^i_\mp + \frac{\i}{2} \gamma_\mu \eta^i_\mp\ ,
\ee
or we satisfy the integrability conditions that follow from squaring the variations. Here we resort to the latter strategy, using that the commutator of two covariant derivatives from the definition  \eqref{eq:covderspinor} leads to
\be
	[D_\mu, D_\nu] \epsilon^i_\pm = - \tfrac14 R_{\mu\nu}{}^{\rho \sigma} \gamma_{\rho\sigma} \epsilon^i_\pm + \tfrac12 R(\cV)_{\mu\nu}{}^i{}_j \epsilon^j_\pm\ .
\ee
The right hand side obtained by the successive application of \eqref{eq:gravvar} instead depends crucially on the details of the gauging.

\subsubsection*{No gauging}
In this case $\eta^i_\pm = 0$ and we find
\be
	[D_\mu, D_\nu] \epsilon^i_\pm  =  \frac{\i}{8} (\nabla_\mu T^\mp_{\nu \rho}) \gamma^\rho \epsilon^i_\mp +  \frac{1}{64} T^\mp_{\mu\rho} \gamma^\rho T^\pm_{\nu\sigma} \gamma^\sigma \epsilon^i_\pm  - (\mu \leftrightarrow \nu)\ .
\ee

\subsubsection*{FI gauging}
Using \eqref{eq:s-gauged} we find
\bea
\begin{split}
	[D_\mu, D_\nu] \epsilon^i_\pm  &= [ \frac{\i}{8} (\nabla_\mu T^\mp_{\nu \rho}) \gamma^\rho \epsilon^i_\mp + \frac{1}{64}  T^\mp_{\mu\rho} \gamma^\rho T^\pm_{\nu\sigma} \gamma^\sigma \epsilon^i_\pm - (\mu \leftrightarrow \nu)] \\
& - \frac1{2 L^2} \gamma_{\mu\nu} \epsilon^i_\pm +\frac{\i}{16 L} [ T^\mp_{\mu\rho} \gamma^{\rho} \gamma_{\nu} + T^\pm_{\nu\rho} \gamma_\mu \gamma^{\rho} - (\mu \leftrightarrow \nu)] \sigma_3{}^i{}_j \epsilon^j_\pm\ .
\end{split}
\eea

\subsection{Omega deformations}
\label{subsec:omegas}
Thus far we listed the requirements on the bosonic fields following solely from the requirement of independent $Q$-variations, as far as this could be pushed (we have kept some generality in order to also allow for the half-BPS $\Omega\, \mathbb{H}^4$, see below). This is however not enough for us to zoom in on the configurations of interest, since we know that other backgrounds such as $\mathbb{H}^2 \times$S$^2$ and the pp-wave will solve the same equations, see \cite{Hristov:2009uj}. Therefore we now specify our interest only in a metric of maximally symmetric space, and allow for additional deformations from purely (anti-) self-dual two-forms:
\be
	 F^{I, +} \sim T^+ = 0\ , \qquad \text{or} \qquad F^{I, -} \sim T^- = 0\ .
\ee
For concreteness (and without loss of generality due to the obvious symmetry) we pick the first option and only allow for non-vanishing $T^-$ and $F^{I, -}$.

\subsubsection*{No gauging: the Nekrasov-Okounkov background}
In the lack of gauging we simply have 
\be
	R_{\mu \nu \rho \sigma} = 0\ ,
\ee
which also requires
\be
	\nabla_\mu T^-_{\nu \rho} = 0\ ,
\ee 
such that all remaining BPS variations are satisfied in the simplest possible way, i.e.\ by the vanishing of each individual term.

The unique background corresponding to these two conditions is the Nekrasov-Okounkov deformation on flat space, $\Omega\, \mathbb{R}^4$, given in Cartesian coordinates as:
\be
	{\rm d} s^2 = {\rm d} x_1^2 + {\rm d} x_2^2 + {\rm d} x_3^2 + {\rm d} x_4^2\ ,
\ee
and
\be
	T^- = 2 (\epsilon_1 - \epsilon_2)\, ({\rm d} x_1 \wedge {\rm d} x_2 - {\rm d} x_3 \wedge {\rm d} x_4)\ , 
\ee
such that the gauge fields are given by
\be
	W^I = \tfrac{X^I}4\,   (\epsilon_1 - \epsilon_2)\, (x_{[1}  {\rm d} x_{2 ]} - x_{[3}  {\rm d} x_{4 ]})\ .
\ee
It is instructive for later purposes to change the Cartesian coordinates on each copy of $\mathbb{R}^2$ to polar, $x_{1, 3} = \rho_{1, 2} \cos \varphi_{1,2}$, $x_{2,4} = \rho_{1,2} \sin \varphi_{1,2}$:
\be
\label{eq:twopolars}
	{\rm d} s^2 = {\rm d} \rho_1^2 + \rho_1^2\, {\rm d} \varphi_1^2 + {\rm d} \rho_2^2 + \rho_2^2\, {\rm d} \varphi_2^2\ ,
\ee
with
\be
	T^- = 2 (\epsilon_1 - \epsilon_2)\, (\rho_1\, {\rm d} \rho_1 \wedge {\rm d} \varphi_1  - \rho_2\, {\rm d} \rho_2 \wedge {\rm d} \varphi_2)\ , 
\ee
and
\be
	W^I = \tfrac{X^I}8\,   (\epsilon_1 - \epsilon_2)\, (\rho_1^2\,  {\rm d} \varphi_1 - \rho_2^2\,  {\rm d} \varphi_2)\ .
\ee
In turn we can convert to spherical Hopf coordinates, $\rho_1 = r \cos \theta$, $\rho_2 = r \sin \theta$:
\be
\label{eq:Hopfcoord}
	{\rm d} s^2 = {\rm d} r^2 + r^2 \left( {\rm d} \theta^2 + \cos^2 \theta\, {\rm d} \varphi_1^2 + \sin^2 \theta\, {\rm d} \varphi_2^2 \right) \ ,
\ee
as well as
\bea
\begin{split}
	T^- &=& 2 (\epsilon_1 - \epsilon_2)\, \Big( r \cos^2 \theta\,  {\rm d} r \wedge {\rm d}  \varphi_1 - r^2 \sin \theta \cos \theta\,  {\rm d} \theta \wedge {\rm d}  \varphi_1 \\ 
&&- r \sin^2 \theta\,  {\rm d} r \wedge {\rm d}  \varphi_2 - r^2 \sin \theta \cos \theta\,  {\rm d} \theta \wedge {\rm d}  \varphi_2  \Big)\ , 
\end{split}
\eea
and
\be
	W^I = \tfrac{X^I}4\,   (\epsilon_1 - \epsilon_2)\,  (r^2  \cos^2 \theta\, {\rm d} \varphi_1 - r^2 \sin^2 \theta\,  {\rm d} \varphi_2)\ .
\ee
This form of $\Omega\, \mathbb{R}^4$ background, as also presented in the introduction, allows for a direct comparison with the $\Omega\, \mathbb{H}^4$ background to which we turn next.

\subsubsection*{FI gauging: the Martelli-Passias-Sparks background}
In the presence of gauging we find the hyperbolic space with 
\be
	R_{\mu\nu\rho\sigma} = \frac1{L^2} \left( g_{\mu \rho} g_{\nu \sigma} - g_{\mu \sigma} g_{\nu \rho} \right)\ ,
\ee
leading to a constant positive curvature~\footnote{In the present conventions, which differ in this respect to most of the other literature, the hyperbolic space (as well as anti-de Sitter space) exhibits constant positive curvatuire.}
\be
	R_{\mu\nu} = \frac3{L^2}  g_{\mu \nu}\ , \qquad R = \frac{12}{L^2}\ ,
\ee
with a metric in spherical Hopf coordinates
\be
	{\rm d} s^2 = \frac{{\rm d} r^2}{1+\tfrac{r^2}{L^2}} + r^2 \left( {\rm d} \theta^2 + \cos^2 \theta\, {\rm d} \varphi_1^2 + \sin^2 \theta\, {\rm d} \varphi_2^2 \right) \ .
\ee
The remaining equation stemming from the Killing spinor integrability condition in this case is non-trivial, 
\be
\label{eq:mpstensor}
\begin{split}
	R(\cV)_{\mu\nu}{}^i{}_j \epsilon^j_\pm = &   [ \frac{\i}{4} (\nabla_\mu T^\mp_{\nu \rho}) \gamma^\rho \epsilon^i_\mp - (\mu \leftrightarrow \nu)] \\
	& +\frac{\i}{8 L} [ T^\mp_{\mu\rho} \gamma^{\rho} \gamma_{\nu} + T^\pm_{\nu\rho} \gamma_\mu \gamma^{\rho} - (\mu \leftrightarrow \nu)] \sigma_3{}^i{}_j \epsilon^j_\pm\ .
\end{split}
\ee
We can further use our assumption that $T^+ = 0$ to find
\be
	R(\cV)_{\mu\nu}{}^i{}_j = -\frac{\i}{4 L} T^-_{\mu\nu}\, \sigma_3{}^i{}_j\ ,
\ee
such that the two different chiralities in \eqref{eq:mpstensor} produce two rather different equations. The lower chirality gives
\be
	T^-_{\mu\nu}\, \sigma_3{}^i{}_j \epsilon^j_- = \tfrac12 [ T^-_{\mu\rho} \gamma_\nu \gamma^{\rho} - T^-_{\nu\rho} \gamma_\mu \gamma^{\rho}] \sigma_3{}^i{}_j \epsilon^j_-\ ,
\ee
while the upper chirality gives
\be
\label{eq:upperchirality}
	T^-_{\mu\nu}\, \sigma_3{}^i{}_j \epsilon^j_+ = L\, \nabla_\rho T^-_{\mu \nu} \gamma^\rho \epsilon^i_- -\tfrac12 [T^-_{\mu\rho} \gamma^{\rho} \gamma_\nu  - T^-_{\nu\rho} \gamma^{\rho} \gamma_\mu] \sigma_3{}^i{}_j \epsilon^j_+\ ,
\ee
where we used the Bianchi identity for $T^-$ (we already know it is an exact form from the relation \eqref{eq:genFtoT}). The first equation is identically satisfied using the anti-self-duality of $T^-$ together with the chirality of the spinor, without any further conditions on $\epsilon_-$. The second equation is however non-trivial, and as shown explicitly in the original reference, \cite{Martelli:2011fu}, one can solve it uniquely by fixing the degrees of freedom of $\epsilon_+$ entirely in terms of $\epsilon_-$, thus reducing by half the amount of supersymmetry generators. The non-vanishing components of the tensor field then take the form
\bea
	\begin{split}
	T^-_{\theta \varphi_1} & = - \frac{2 (b_1^2-b_2^2) (b_1 + b_2 \sqrt{1+\tfrac{r^2}{L^2}})\, r^2 \sin \theta \cos \theta}{L\, \Xi^{3/2} (r, \theta)} = \frac{ (b_1 + b_2 \sqrt{1+\tfrac{r^2}{L^2}}) }{(b_2 + b_1 \sqrt{1+\tfrac{r^2}{L^2}})}  T^-_{\theta \varphi_2} \ , \\
	T^-_{r \varphi_1} & =  \frac{2 (b_1^2-b_2^2) (b_2 + b_1 \sqrt{1+\tfrac{r^2}{L^2}})\, r \cos^2 \theta}{L\, \sqrt{1+\tfrac{r^2}{L^2}}\, \Xi^{3/2} (r, \theta)}  = - \frac{ (b_2 + b_1 \sqrt{1+\tfrac{r^2}{L^2}}) \cos^2 \theta}{(b_1 + b_2 \sqrt{1+\tfrac{r^2}{L^2}}) \sin^2 \theta}  T^-_{r \varphi_2} \ , 
\end{split}
\eea
where for brevity we defined
\be
\label{eq:Xirtheta}
	\Xi(r, \theta) :=(b_1 + b_2 \sqrt{1+\tfrac{r^2}{L^2}})^2 \sin^2 \theta +  (b_2 + b_1 \sqrt{1+\tfrac{r^2}{L^2}})^2 \cos^2 \theta\ ,
\ee
leading in turn to the background gauge fields
\be
	W^I = -L\, X^I \frac{(b_1 + b_2 \sqrt{1+\tfrac{r^2}{L^2}})\, {\rm d} \varphi_1 +  (b_2 + b_1 \sqrt{1+\tfrac{r^2}{L^2}})\, {\rm d} \varphi_2}{2\, \sqrt{\Xi(r, \theta) }}\ ,
\ee
as already anticipated. It is then easy to check that $\nabla^\mu F^{I, -}_{\mu \nu} = \nabla^\mu T^-_{\mu \nu} = 0$. We thus arrive at the following relation between the Killing spinors of opposite chiralities,
\be
\begin{split}
	\epsilon^i_+ = & \frac{r}{L\, \Xi(r, \theta)}\, \Big[ (b_1^2 - b_2^2) \sqrt{1+\tfrac{r^2}{L^2}} \sin \theta \cos \theta\, \gamma^1 \\
	& - \left( b_1 b_2 \sqrt{1+\tfrac{r^2}{L^2}} + b_1^2 \sin^2 \theta + b_2^2 \cos^2 \theta  \right) \gamma^2 \Big]\, \sigma_3{}^i{}_j \epsilon^j_-\ . 
\end{split}
\ee
We are not going to need here the explicit solution for $\epsilon_\pm$, see \cite{Martelli:2011fu,Farquet:2014kma}, but we reproduce for completeness the fact that the Killing spinor bilinears produce the so-called {\it canonical} isometry $\xi$ with a fixed point at the centre of $\mathbb{H}^4$ at $r=0$, where locally
\be
\label{eq:xilocal}
	\xi = b_1\, \partial_{\varphi_1} + b_2\, \partial_{\varphi_2}\ ,
\ee
as discussed at length in \cite{BenettiGenolini:2019jdz}.

Finally, contracting \eqref{eq:upperchirality} with $\gamma^{\mu\nu}$ leads to the identity
\be
	\gamma^{\mu\nu} \slashed{\nabla} T^-_{\mu\nu} \epsilon^i_- = 0\ ,
\ee
such that the last remaining supersymmetry constraint, the first equation in \eqref{eq:dilatinosolve}, is satisfied.

Note that the explicit Killing spinors constructed in \cite{Martelli:2011fu} not only vary from those on $\mathbb{H}^4$ in the number of free parameters, but also in their functional dependence. In the $\Omega\, \mathbb{H}^4$ case the Killing spinors are only functions of $r$ and $\theta$ and are independent of the $\varphi_1$ and $\varphi_2$ coordinates. We come back to this remark at the end of the paper when discussing conical defects.

\subsection{Summary of off-shell configurations}
\label{subsec:summary}
The off-shell backgrounds can be briefly summarized as follows for the benefit of the readers uninterested in the technical derivation.

\subsubsection*{$\Omega\, \mathbb{R}^4$}
\be
\begin{split}
	{\rm d} s^2 & = \sum_{i=1}^4 {\rm d} x_i^2\ , \qquad X^I_\pm  = const\ , \qquad  F^{I, \pm}_{\mu\nu} = \frac14 X^I_\pm T^\pm_{\mu\nu}\ ,   \\
		D & = 0\ , \qquad \cV_\mu {}^i {}_j = 0\ , \qquad \chi_H^{-1/2} A_i{}^\alpha = \delta_i{}^\alpha\ , \qquad Y^{I, ij} = 0\ , \\
	   T^+_{\mu\nu} & = 0\ , \qquad  T^-_{12} = (\epsilon_1 - \epsilon_2)\, \qquad T^-_{13} = T^-_{14} = 0\ .
\end{split}
\ee

\subsubsection*{$\Omega\, \mathbb{H}^4$}
\be
\label{eq:fullOH}
\begin{split}
	{\rm d} s^2 & =  \frac{{\rm d} r^2}{1+\tfrac{r^2}{L^2}} + r^2 \left( {\rm d} \theta^2 + \cos^2 \theta\, {\rm d} \varphi_1^2 + \sin^2 \theta\, {\rm d} \varphi_2^2 \right) \ , \qquad D  = 0\ ,  \\
		 \cV_\mu {}^i {}_j &= -\i\, g_I W^I_\mu \sigma_3{}^i{}_j\ , \qquad \chi_H^{-1/2} A_i{}^\alpha = \delta_i{}^\alpha\ , \qquad Y^I_{ij} = \frac{2\, \i}{L} X^I \varepsilon_{i k} \sigma_3{}^k{}_j\ , \\
 X^I_+ & = X^I_-  = X^I= const\ , \qquad g_I X^I = \frac1{L}\ , \qquad F^{I, \pm}_{\mu\nu} = \frac14 X^I_\pm T^\pm_{\mu\nu}\ ,   \\
	   T^+_{\mu\nu} & = 0\ , \qquad  T^-_{\theta \varphi_1} = - \frac{2 (b_1^2-b_2^2) (b_1 + b_2 \sqrt{1+\tfrac{r^2}{L^2}})\, r^2 \sin \theta \cos \theta}{L\, \Xi^{3/2} (r, \theta)}\ , \\
T^-_{r \theta} & = 0\ , \qquad T^-_{r \varphi_1} =  \frac{2 (b_1^2-b_2^2) (b_2 + b_1 \sqrt{1+\tfrac{r^2}{L^2}})\, r \cos^2 \theta}{L\, \sqrt{1+\tfrac{r^2}{L^2}}\, \Xi^{3/2} (r, \theta)}\ .
\end{split}
\ee

\section{Off-shell action}
\label{sec:off}
So far we introduced all fundamental fields in the superconformal formalism and analysed the supersymmetry variations of the fermionic fields without the need to write down a specific Lagrangian for the theory. It is an important consequence of realizing supersymmetry in an off-shell way that BPS configurations do not depend on the explicit theory, with the important exception of the gauging choice that we discussed in the hypermultiplet sector. In order to explicitly evaluate the corresponding action and in turn look at physical observables on a given background, it is however indispensable that 
we commit to a particular theory. As announced in the introduction, we are interested in higher derivative supergravity with two infinite towers of higher derivative terms labeled by $\mathbb{W}$ and $\mathbb{T}$. The theory of interest coincides with the one in \cite{Hristov:2021qsw}, but here we stick to Euclidean signature. Similarly to many of the fundamental fields we discussed above, this also means there is a natural split into a chiral and an anti-chiral part of the composite multiplets needed for the introduction of the higher derivative terms, see again \cite{deWit:2017cle} and \cite{Bobev:2021oku},
\be
	\Phi^p_\pm =  \{A^\pm_p, \Psi^{i, p}_\pm, \cB^{ij}_{p, \pm}, \cG^\pm_{p, \mu\nu}, \Lambda^{i, p}_\pm, C_{p, \pm}\}\ ,
\ee
where the index  $p \in \{\mathbb{W}, \mathbb{T} \}$ denotes the multiplet in question. These composite multiplets can be uniquely specified by defining their lowest components, $A^\pm_p$, in terms of the fundamental fields, see below. We can then specify uniquely the Lagrangian by the choice of gauging discussed earlier, together with the choice of prepotentials
\be
\label{eq:prepots}
	F^\pm (X^I_\pm; A^\pm_\mathbb{W},  A^\pm_\mathbb{T}) := \sum_{m, n = 0}^\infty \, F_\pm^{(m,n)} (X^I_\pm)\, (A^\pm_\mathbb{W})^m\, (A^\pm_\mathbb{T})^n\ ,  
\ee
whose chiral and anti-chiral components are in principle now completely independent. As usual also in Poincar\'e supergravity, derivatives of $F$ w.r.t.\ $X^I$ are denoted by $F_I$ and w.r.t.\ $A_p$ by $F_{A_p}$. The Lagrangian is then given by 
\begin{equation}
\label{eq:Eucl-action}
\begin{split}
e^{-1}\mathcal{L} &\; =   e^{-\cK} (\tfrac16 R - D) - D_\mu F_I^+ D^\mu X^I_-  - D_\mu F_I^- D^\mu X^I_+ + \tfrac{1}{32}F^+\bigl(T_{ab}^+)^2 + \tfrac1{32}F^-\bigl(T_{ab}^-\bigr)^2 \\
&\;-\tfrac14\,F_{IJ}^+\,\cF_{ab}^{-I}\cF^{ab-J} - \tfrac14\,F_{IJ}^-\,\cF_{ab}^{+I}\cF^{ab+J} + \tfrac18\,F_I^+\,\cF_{ab}^{+I} T^{ab+} + \tfrac18\,F_I^-\,\cF_{ab}^{-I} T^{ab-}   \\
&\;+\tfrac18\, N_{IJ}\,Y_{ij}^I Y^{J, ij} + \tfrac12\,F^+_{A_p}\,\mathcal{C}_{p, +} + \tfrac12\,F^-_{A_p}\,\mathcal{C}_{p, -} + \tfrac14\,F^+_{A_p I}\,\mathcal{B}^{ij}_{p, +}\,Y_{ij}^I + \tfrac14\,F^-_{A_p I}\,\mathcal{B}^{ij}_{p, -}\,Y_{ij}^I \\
&\;- \tfrac12\,F^+_{A_p I}\,\cG_{p, ab}^-\,\cF^{ab-I} - \tfrac12\,F^-_{A_p I}\,\cG_{p, ab}^+\,\cF^{ab+I} + \tfrac18\,F^+_{A_p A_q}\,\mathcal{B}_{p, ij+}\mathcal{B}^{ij}_{q, +} \\ 
&\;+ \tfrac18\,F^-_{A_p A_q}\,\mathcal{B}_{p, ij-}\mathcal{B}^{ij}_{q, -} - \tfrac14\,F^+_{A_p A_q}\,\cG_{p, ab}^-\cG^{ab -}_q - \tfrac14\,F^-_{A_p A_q}\,\cG_{p, ab}^+\cG^{ab+}_q \, \\
&\; +\chi_\mathrm{H} (\tfrac16 R + \tfrac12 D ) - \tfrac12\,\varepsilon^{ij}\Omega_{\alpha\beta}\, D_\mu A_i{}^\alpha D^\mu A_j{}^\beta \\
&\;+\tfrac12\,\varepsilon^{ij}\Omega_{\alpha\beta}\,A_i{}^\alpha\,(g_I X^I_+)(g_J X^J_-)\,t^\beta{}_\gamma t^\gamma{}_\delta\,A_j{}^\delta - \tfrac14\,\Omega_{\alpha\beta}\,A_i{}^\alpha\,(g_I Y^{ijI})\,t^\beta{}_\gamma\,A_j{}^\gamma\, ,
\end{split}
\end{equation}  
where
\be
	e^{-\cK} := F^+_I X^I_- + F^-_I X^I_+\ , \qquad  N_{IJ} := F^+_{IJ}+F^-_{IJ}\ ,
\ee
and
\be
	\cF^{\pm I}_{ab} := F^{\pm I}_{ab} -\tfrac14 X^I_\pm T^\pm_{ab}\ ,
\ee
which actually vanishes identically on the backgrounds we consider here. Note that the last two rows of the Lagrangian above come from the auxiliary hypermultiplet, $e^{-1} \cL_\mathrm{H}$. 

At this stage we can start simplifying our Lagrangian for the purpose of evaluating the action on the backgrounds of interest. Our first general assumption is that the theory exhibits a {\it real} Lagrangian and action, which means
\be
	F_+^{(m,n)} (X^I) = F^{(m,n)}_- (X^I) = F^{(m,n)} (X^I)\ ,
\ee
as also discussed in \cite{Bobev:2021oku}. Note that this {\it does not} correspond to an equality of the full prepotentials $F_+$ and $F_-$ as defined in \eqref{eq:prepots}, which depend on a set of a priori independent fields. We are thus going to keep the subscripts until the final evaluation where confusion is no longer possible.

Let us now observe that actually many of the terms in the Lagrangian conveniently vanish on the backgrounds of interest, even before we go into the details of the higher derivative terms. The hypermultiplet Lagrangian, using \eqref{eq:FIgauge} and \eqref{eq:gaugefix}, becomes
\be
	e^{-1} \cL_\mathrm{H} = \chi_\mathrm{H} (\tfrac16 R + \tfrac12 D - 2 (g_I X^I)^2)\ ,
\ee
which vanishes both the the $\Omega\, \mathbb{R}^4$ and the $\Omega\, \mathbb{H}^4$ backgrounds we consider (as well as on the corresponding undeformed vacua). Usually the hypermultiplet Lagrangian is kept in order to gauge-fix the hyper-K\"ahler potential $\chi_\mathrm{H}$ and introduce the Newton constant in the on-shell action via the $D$ field equation of motion. Here we take a different path to introducing $\kappa$ (see section \ref{sec:onshell}) and therefore have no need for the equation of motion for the scalar $D$, which itself is vanishing on the backgrounds of interest. Together with the observation that $\cF^{\pm I}_{ab} = 0$ and $X^I_+ = X^I_- = const$ on all backgrounds of interest, we arrive at a substantially simpler off-shell action on maximally symmetric backgrounds,
\begin{equation}
\label{eq:Eucl-action-max}
\begin{split}
e^{-1}\mathcal{L} &\; =\tfrac16\, e^{-\cK}\,  R\,  + \tfrac1{32}\,F^- \bigl(T_{ab}^-\bigr)^2 + \tfrac18\, N_{IJ}\,Y_{ij}^I Y^{J, ij}\, \\
&\; + \tfrac12\,F^+_{A_p}\,\mathcal{C}_{p, +} + \tfrac12\,F^-_{A_p}\,\mathcal{C}_{p, -} + \tfrac14\,F^+_{A_p I}\,\mathcal{B}^{ij}_{p, +}\,Y_{ij}^I + \tfrac14\,F^-_{A_p I}\,\mathcal{B}^{ij}_{p, -}\,Y_{ij}^I \\
&\;+ \tfrac18\,F^+_{A_p A_q}\,\mathcal{B}_{p, ij+}\mathcal{B}^{ij}_{q, +} + \tfrac18\,F^-_{A_p A_q}\,\mathcal{B}_{p, ij-}\mathcal{B}^{ij}_{q, -}  \\ 
&\;- \tfrac14\,F^+_{A_p A_q}\,\cG_{p, ab}^-\cG^{ab -}_q - \tfrac14\,F^-_{A_p A_q}\,\cG_{p, ab}^+\cG^{ab+}_q \, ,
\end{split}
\end{equation}  
where we also implemented the vanishing of the self-dual tensor $T^+$.

At this stage we need to finally define properly the $\mathbb{W}$ and $\mathbb{T}$ multiplets in order to start simplifying the action further. The $\mathbb{W}$ invariant is defined by the Weyl multiplet written in a chiral/anti-chiral form
\be
	A^\pm_\mathbb{W} = \tfrac1{64}(T_{ab}^\mp)^2\ , 
\ee
and in turn the remaining bosonic components of the corresponding chiral and anti-chiral multiplets can be evaluated to
\be
\label{eq:W-mult}
\begin{split}
	B^{ij}_{\mathbb{W}, \pm} & = \tfrac14 T^{\mp ab} \varepsilon^{k ( i} R(\cV)_{ab}{}^{j)}{}_k\ , \\
	\cG_{\mathbb{W}, a b}^\pm & = 0\ , \\
	\cC_{\mathbb{W}, \pm} & = - \tfrac12 ( R(\cV)^{\mp}_{ab}{}^i{}_j )^2\ .
\end{split}
\ee
Note that we did not write down the most general expressions that are readily available in \cite{deWit:2017cle}, but directly simplified them on the backgrounds of interest. In particular we omitted a term proportional to $D^a T^-_{a b}$ from the last equality and simplified the remaining expressions using that the dilatational and $\SO(1,1)$ gauge fields vanish and $D = 0$. We further used the vanishing of the Weyl tensor on $\mathbb{R}^4$ and $\mathbb{H}^4$.

The chiral multiplet related to the $\mathbb{T}$ invariant is chosen to be made out of the linear combination of the (anti-) chiral multiplets $g_I \cX^I_\pm$ that make up the vector multiplets (in the ungauged case all components are identically zero for an arbitrary choice of $\mathbb{T}$ invariant), 
\be
	 A^\pm_\mathbb{T} = \frac{\Box_\mathrm{c} (g_I X^I_\pm)}{(g_J X^J_\mp)} + \tfrac18 \frac{g_I \cF^{I, \mp}_{ab} T^{\mp, ab}}{(g_J X^J_\mp)} - \frac{g_I g_J}{8 (g_K X^K_\mp)} \left( Y^I_{ij} Y^{J, ij} - 2 \cF^{I, \pm}_{ab} \cF^{J, \pm ab} \right)\ ,
 \ee
see \cite{Butter:2013lta} and \cite{Butter:2014iwa} for more details. Using that the scalars $X^I_+ = X^I_- = X^I$ and $Y^I_{ij}$ are constant and $\cF^{\pm I}_{ab} = 0$, we can simplify the general expressions for the bosonic fields in the resulting multiplet (see also \cite{Banerjee:2016qvj} for more explicit formulae),
\be
\label{eq:W-mult}
\begin{split}
	 A^\pm_\mathbb{T} & = \frac16 R- \frac{(g_I Y^I_{ij}) (g_J Y^{J, ij})}{8 (g_K X^K)^2}  \ , \\
	B^{ij}_{\mathbb{T}, \pm} & = \frac16 R \frac{g_I Y^{I, ij}}{g_J X^J}  \ , \\
	\cG_{\mathbb{T}, a b}^\pm & =  R(\cV)^\pm_{ab}{}^i{}_k \frac{g_I Y^{I, jk}}{g_J X^J} \varepsilon_{i j} - \frac{R}{12} T^\pm_{ab} \ , \\
	\cC_{\mathbb{T}, \pm} & = \tfrac13 R^2 - (R_{ab})^2 - \tfrac12 (R(\cV)^\pm_{ab}{}^i{}_j)^2 - \tfrac1{16} (T^\pm_{ab})^2 A^\pm_\mathbb{T}\ .
\end{split}
\ee

We have simplified the Lagrangian and the higher derivative corrections as much as possible, and now we are in a position to evaluate the resulting action on the particular BPS configurations of interest.

\subsection{$\Omega\, \mathbb{R}^4$}
\label{subsec:31}
Using the explicit background configuration of $\Omega\, \mathbb{R}^4$, it is immediate to find that the entire $\mathbb{T}$ multiplet vanishes identically,
\be
	 A^\pm_\mathbb{T} =  B^{ij}_{\mathbb{T}, \pm} = \cG_{\mathbb{T}, a b}^\pm   = \cC_{\mathbb{T}, \pm} = 0\ ,
\ee
as similarly observed in \cite{Butter:2014iwa} for asymptotically flat black holes, while the $\mathbb{W}$ multiplet takes the simple form
\be
	 A^+_\mathbb{W} = \tfrac1{16} (\epsilon_1 - \epsilon_2)^2\ , \quad A^-_\mathbb{W}= B^{ij}_{\mathbb{W}, \pm} = \cG_{\mathbb{W}, a b}^\pm   = \cC_{\mathbb{W}, \pm} = 0\ .
\ee
We therefore find two different expressions for the prepotentials $F^+$ and $F^-$,
\be
	F^- = F^{(0,0)}\ , \qquad F^+ =  \sum_{m=0}^\infty F^{(m,0)}\,  \frac{(\epsilon_1 - \epsilon_2)^{2 m}}{4^m}\ .
\ee

Notice that we actually have only a single non-vanishing term in the Lagrangian \eqref{eq:Eucl-action-max}, proportional to $F^- (T^-)^2$, meaning that {\it all} higher derivative contributions are set to zero as they only feature inside $F^+$ above. We thus find
\be
	e^{-1} \cL |_ {\Omega\, \mathbb{R}^4} = \frac18 (\epsilon_1 - \epsilon_2)^2\, F^{(0,0)} (X^I)\ .
\ee 

Let us now discuss the evaluation of the action, noting that we need to add the usual Gibbons-Hawking-York \cite{Gibbons:1976ue,York:1986it} boundary term in order to have a well-defined action principle,
\be
	S_\text{bdry} = - \int_{\partial M} {\rm d} x^3\, \sqrt{h}\, \frac{(e^{-\cK} + \chi_\mathrm{H})}{3} \left( K - K^0 \right)\ , 
\ee
where $h_{ab}$ is the induced metric on the boundary (a three-sphere at infinity in the coordinates \eqref{eq:Hopfcoord}), $K_{ab}$ is the extrinsic curvature and $K^0_{ab}$ the additional boundary extrinsic curvature needed to normalize the flat space action to zero,
\be
	S |_ {\mathbb{R}^4} = 0\ .
\ee
In the presence of the deformation it is easy to see that the above boundary term is not enough to regularize the action, which can be formally evaluated using a naive cutoff at radius $r = \Lambda$ that should be sent to infinity:
\be
	S |_ {\Omega\, \mathbb{R}^4} = \lim_{\Lambda \rightarrow \infty} \tfrac{\pi^2}{16}\,  (\epsilon_1 - \epsilon_2)^2\, \Lambda^4\, F^{(0,0)} (X^I)\ .
\ee
Notice that, importantly, the contribution from the centre of the space (the position of the nut) vanishes. This infinite asymptotic contribution can easily be regularized by an additional boundary term proportional to $\int_{\partial M}{\rm d}^3 \Sigma^\mu\, T^-_{\mu\nu}\, W^{I, \nu}$, see \cite{Braden:1990hw}, but this immediately leads to a vanishing action and does not have the clear physical interpretation of changing the ensemble as in \cite{Braden:1990hw}.

\subsection{$\mathbb{H}^4$}
In order to ease our way into the full $\Omega\, \mathbb{H}^4$ calculation which is considerably more involved, we first focus on the case of pure $\mathbb{H}^4$ in the lack of a deformation. We thus turn off the two-form field $T^-$ and in turn have vanishing gauge fields $W^I_\mu = 0$ and $\SU(2)$ field $\cV_\mu{}^i{}_j = 0$. The $\mathbb{W}$ multiplet fields then vanish entirely,
\be
	 A^\pm_\mathbb{W} =  B^{ij}_{\mathbb{W}, \pm} = \cG_{\mathbb{W}, a b}^\pm   = \cC_{\mathbb{W}, \pm} = 0\ ,
\ee
and, using $R = 12/L^2$, the $\mathbb{T}$ fields are given by
\be
 A^\pm_\mathbb{T}   = \tfrac1{L^2}\ , \qquad   B^{ij}_{\mathbb{T}, \pm} = - \tfrac{4 \i}{L^3} \varepsilon^{j k} \sigma_3{}^i{}_k \ , \qquad \cG_{\mathbb{T}, a b}^\pm = 0\ , \qquad \cC_{\mathbb{T}, \pm} = \tfrac{12}{L^4}\ . 
\ee
In order to evaluate the Lagrangian \eqref{eq:Eucl-action-max}, we can also simplify the prepotentials and their derivatives,
\be
\begin{split}
	F^\pm & = \sum_{n=0}^\infty F^{(0,n)}\,  L^{- 2 n}\ , \quad F^\pm_{I J} = \sum_{n=0}^\infty F_{I J}^{(0,n)}\,  L^{- 2 n}\ , \quad F^\pm_{A_\mathbb{T}} = \sum_{n=0}^\infty F^{(0,n)}\, n L^{2 (1- n)}\ , \\ 
 F^\pm_{I A_\mathbb{T}} & = \sum_{n=0}^\infty F_I^{(0,n)}\, n L^{2 (1- n)}\ , \qquad \qquad F^\pm_{A_\mathbb{T} A_\mathbb{T}} = \sum_{n=0}^\infty F^{(0,n)}\, n (n-1) L^{2 (2- n)}\ . 
\end{split}
\ee
We also remind the reader that the homogeneity properties of the prepotentials imply
\be
\begin{split}
	X^I F_I^{(0,n)} & = 2 (1-n) F^{(0,n)}\ , \qquad X^J F_{IJ}^{(0,n)} = (1- 2 n)  F_I^{(0,n)}\ , \\
	& \Rightarrow X^I X^J F_{IJ}^{(0,n)} = 2 (n-1) (2 n-1) F^{(0,n)}\ .
\end{split}
\ee
We also need the simplifying identities for the scalar triplets,
\be
\label{eq:Bid}
	B^{ij}_{\mathbb{T}, \pm} Y^I_{ij} = \tfrac{16}{L^4}\, X^I\ , \qquad B^{ij}_{\mathbb{T}, \pm} B_{ij \mathbb{T}, \pm} = \tfrac{32}{L^6}\ ,
\ee
derived using \eqref{eq:simplifY}.

Plugging all these identities in the off-shell Lagrangian produces a remarkable simplification,
\be
\label{eq:Hlag}
\begin{split}
	e^{-1} \cL |_{\mathbb{H}^4} = & \frac{4}{L^2} \sum_{n=0}^\infty \frac{F^{(0,n)}}{L^{2 n}} \Big[ 2 (1-n) + (1-n) (1-2 n)  + 3n \\
& +4n (1-n) + 2 n (n-1) \Big] = \frac{12}{L^2} \sum_{n=0}^\infty \frac{F^{(0,n)}}{L^{2 n}} = \frac{3}{L^2}\, F(2 X^I; 0, 4 L^{-2})  \ .
\end{split}
\ee

The final step of evaluating the (in this case {\it off-shell}) action passes through the procedure of {\it holographic renormalization}, see \cite{Skenderis:2002wp}. The general approach to this task, when applied to arbitrary asymptotically Euclidean AdS configurations, involves an off-shell generalization of this procedure in the presence of arbitrary higher derivative corrections, notably involving terms that are quadratic in the Riemann curvature. A useful discussion on this topic was presented in \cite{Bobev:2021oku} restricted to the four derivative couplings, and it would be desirable to understand this procedure in the presence of scalar couplings and arbitrary number of derivatives. Here we have a much simpler task of evaluating the action on the empty vacuum, where we have simplified the Lagrangian as above. It is therefore clear that we can simply add to the off-shell action above the usual two derivative Gibbons-Hawking-York boundary term,  \cite{Gibbons:1976ue,York:1986it}, with the suitable prefactor, 
\be
\label{eq:Hct}
	S_\text{bdry} = - \int_{\partial M} {\rm d} x^3 \sqrt{h}\, F(2 X^I; 0, 4 L^{-2})  \left( K - \tfrac{L}{2}\, \cR - \tfrac2{L} \right)\ , 
\ee
where $h_{ab}$ is the induced metric on the boundary, $K_{ab}$ is the extrinsic curvature, and $\cR_{abcd}$ is the Riemann tensor of the induced metric. We then find
\be
\label{eq:Hact}
	S  |_{\mathbb{H}^4} = 4 \pi^2 L^2\, F(2 X^I; 0, 4 L^{-2}) =  4 \pi^2\, F(2 L\, X^I; 0, 4)  \ .
\ee	
We see that the Lagrangian \eqref{eq:Hlag} leads straightforwardly to \eqref{eq:Hact} using the counterterms \eqref{eq:Hct} due to the fact that we have fixed all scalar fields to constants. We give more comments on the general procedure of holographic renormalization in the next section when discussing the proof of \cite{Hristov:2021qsw}.

\subsection{$\Omega\, \mathbb{H}^4$}
Now we include the deformation coming from the non-vanishing two-form field, $T^-$. The evaluation of the higher derivative action on the $\Omega\, \mathbb{H}^4$ background now requires the identity
\be
\label{eq:Tsquared}
	\frac{(T^-_{ab})^2}{64} = \frac{ (b^2_1 - b^2_2)^2}{4\, L^2\, \Xi^2 (r, \theta)}\ ,
\ee
using $\Xi(r, \theta)$ as defined in \eqref{eq:Xirtheta}. In order to keep the formulae below more compact, we are going to use the l.h.s.\ expression above until the final step.

Due to the chosen chirality of $T^-$, the anti-chiral part of the $\mathbb{W}$ multiplet fields still vanish. However, the $\mathbb{W}^+$ fields now no longer vanish, and we find
\be
\begin{split}
	A^+_\mathbb{W}  = \tfrac1{64} (T^-_{ab})^2\ , \quad  B^{ij}_{\mathbb{W}, +} & = \tfrac{\i}{16 L} (T^-_{ab})^2 \varepsilon^{j k} \sigma_3{}^i{}_k\ , \quad \cC_{\mathbb{W}, +} =  \tfrac1{16 L^2} (T^-_{ab})^2  \\
	A^-_\mathbb{W} & = B^{ij}_{\mathbb{W}, -} =  \cG_{\mathbb{W}, a b}^\pm = \cC_{\mathbb{W}, -}  =  0\ .
\end{split}
\ee
This natural imbalance between the chiralities reflects into a difference between the prepotentials $F^+$ and $F^-$ that we need to keep track of. The evaluation of the $\mathbb{T}$ multiplet fields instead remains the same as for pure $\mathbb{H}^4$ (after some non-trivial cancellations),
\be
	 A^\pm_\mathbb{T}   = \tfrac1{L^2}\ , \qquad   B^{ij}_{\mathbb{T}, \pm} =  - \tfrac{4 \i}{L^3} \varepsilon^{j k} \sigma_3{}^i{}_k \ ,  \qquad \cG_{\mathbb{T}, a b}^\pm = 0\ , \qquad \cC_{\mathbb{T}, \pm} = \tfrac{12}{L^4}\ .
\ee
In addition to \eqref{eq:simplifY} and \eqref{eq:Bid}, in order to simplify the Lagrangian here we need the following identities
\be
	B^{ij}_{\mathbb{W}, +} Y^I_{ij} = - \tfrac{(T_{ab}^-)^2}{4 L^2}\, X^I\ , \quad B^{ij}_{\mathbb{W}, +} B_{ij \mathbb{T}, \pm} = - \tfrac{(T^-_{ab})^2}{2 L^4}\ , \quad  B^{ij}_{\mathbb{W}, +} B_{ij \mathbb{W}, +} = \tfrac{(T^-_{ab})^4}{128 L^2} \ .
\ee

The relevant prepotentials and their (non-vanishing) derivatives now can be simplified to the form
\be
\begin{split}
	F^+ & = \sum_{m,n=0}^\infty F^{(m,n)}\, \frac{ (T^-_{ab})^{2 m}}{64^m  L^{2 n}}\ , \qquad F^- = \sum_{n=0}^\infty F^{(0,n)}\, \frac1{ L^{2 n}}\ , \\
F_{I J}^+ &= \sum_{m,n=0}^\infty F_{I J}^{(m,n)}\, \frac{ (T^-_{ab})^{2 m}}{64^m  L^{2 n}}\ , \qquad F_{I J}^- = \sum_{n=0}^\infty F_{I J}^{(0,n)}\, \frac1{ L^{2 n}}\ ,  \\
 F^+_{A_\mathbb{T}} & = \sum_{m,n=0}^\infty F^{(m,n)}\, n \frac{ (T^-_{ab})^{2 m}}{64^m  L^{2 (n-1)}}\ , \qquad     F^-_{A_\mathbb{T}}  = \sum_{n=0}^\infty F^{(0,n)}\, n \frac1{L^{2 (n-1)}}\ , \\
F^+_{I A_\mathbb{T}} & = \sum_{m,n=0}^\infty F_I^{(m,n)}\, n \frac{ (T^-_{ab})^{2 m}}{64^m  L^{2 (n-1)}}\ , \qquad F^-_{I A_\mathbb{T}}  = \sum_{n=0}^\infty F_I^{(0,n)}\, n \frac1{L^{2 (n-1)}}\ , \\
F^+_{A_\mathbb{W}} &  = \sum_{m,n=0}^\infty F^{(m,n)}\, m \frac{ (T^-_{ab})^{2 (m-1)}}{64^{(m-1)}  L^{2 n}}\ ,  \qquad
F^+_{I A_\mathbb{W}}  = \sum_{m,n=0}^\infty F_I^{(m,n)}\, m \frac{ (T^-_{ab})^{2 (m-1)}}{64^{(m-1)}  L^{2 n}}\ , \\
 & F^+_{A_\mathbb{T} A_\mathbb{T}}  = \sum_{m,n=0}^\infty F^{(m,n)}\, n (n-1) \frac{ (T^-_{ab})^{2 m}}{64^m  L^{2 (n-2)}}\ , \\
 & F^-_{A_\mathbb{T} A_\mathbb{T}}  = \sum_{n=0}^\infty F^{(0,n)}\, n (n-1) \frac1{ L^{2 (n-2)}}\ , \\
 & F^+_{A_\mathbb{W} A_\mathbb{T}} = \sum_{m,n=0}^\infty F^{(m,n)}\, m n \frac{ (T^-_{ab})^{2 (m-1)}}{64^{(m-1)}  L^{2 (n-1)}}\ , \\
 & F^+_{A_\mathbb{W} A_\mathbb{W}} = \sum_{m,n=0}^\infty F^{(m,n)}\, m (m-1) \frac{ (T^-_{ab})^{2 (m-2)}}{64^{(m-2)}  L^{2 n}}\ ,
\end{split}
\ee
and we again need
\be
\begin{split}
	X^I F_I^{(m,n)} & = 2 (1-m-n) F^{(m,n)}\ , \quad X^J F_{IJ}^{(m,n)} = (1-2 m- 2 n)  F_I^{(m,n)}\ , \\
	& \Rightarrow X^I X^J F_{IJ}^{(m,n)} = 2 (m+n-1) (2 m+ 2 n-1) F^{(m,n)}\ .
\end{split}
\ee
as implied by the homogeneity properties.

The full Lagrangian now features a number of different terms with either a single or a double infinite series, but we can make use of the pure $\mathbb{H}^4$ case above in order to understand their systematics. It turns out that it is convenient to regroup the second series, labeled by the integer $m$, in contributions from $m=0$ and $m > 0$ separately,
\be
\label{eq:OmHlag}
\begin{split}
	e^{-1} \cL |_{\Omega \mathbb{H}^4} =\, & \frac{12}{L^2} \sum_{n=0}^\infty \frac{F^{(0,n)}}{L^{2 n}} + \frac{(T^-_{ab})^2}{32}  \sum_{n=0}^\infty \frac{F^{(0,n)}}{L^{2 n}} + \frac{2}{L^2} \sum_{m=1}^\infty \sum_{n=0}^\infty \frac{F^{(m,n)}}{L^{2 n}} \frac{(T^-_{ab})^{2 m}}{64^m} \Big[ 3 n  \\
& + m +  2 (1 - m - n) +  (1 - m - n) (1 - 2 m - 2n) + 4 n (1 - m - n)  \\
& - 4 m (1-m-n) +2 n (n-1) + 2 m (m-1) - 4 m n  \Big] \\
=\, & \frac{12}{L^2} \sum_{n=0}^\infty \frac{F^{(0,n)}}{L^{2 n}} + \frac{(T^-_{ab})^2}{32}  \sum_{n=0}^\infty \frac{F^{(0,n)}}{L^{2 n}} \\
&+ \frac{2}{L^2} \sum_{m=1}^\infty \sum_{n=0}^\infty \frac{F^{(m,n)}}{L^{2 n}} \frac{(1-2m) (3-4 m) (T^-_{ab})^{2 m}}{64^m}   \ ,
\end{split}
\ee
where it is evident that the first term is the one resulting from pure $\mathbb{H}^4$, which is the reason we did not spell out in more detail how it comes about (see \eqref{eq:Hlag}).

It is now rather interesting to observe that only the first term in the Lagrangian above is constant and requires the addition of counterterms in order to produce a finite action, as already done and discussed around \eqref{eq:Hct}-\eqref{eq:Hact}. All the other terms are proportional to $(T^-)^{2 m}$ and each higher power leads to further subleading terms asymptotically, c.f.\ \eqref{eq:Tsquared}. Therefore, in agreement with holographic renormalization at two derivatives, see \cite{Martelli:2011fu,Farquet:2014kma}, the action is finite without the addition of any boundary terms other than \eqref{eq:Hct}. By a direct integration,~\footnote{We could perform the integral explicitly via {\it Mathematica} for the first five powers in the sequence. The rather non-trivial coefficients predicted correctly give us confidence in the general answer, but it would be interesting to prove this analytically.} we find the remarkable identity
\be
	\frac{2 (1-2m) (3-4 m) }{64^m}\, \int_{\Omega\, \mathbb{H}^4} \sqrt{g}\,  (T^-_{ab})^{2 m} = 4 \pi^2\, L^2\, \frac{(b_1 - b_2)^2}{4 b_1 b_2}\, \frac{(b_1 - b_2)^{2 (m-1)}}{(b_1 + b_2)^{2 (m-1)}}\ ,
\ee
for $m \in \mathbb{Z}_+$, which allows for a seemingly miraculous simplification of the action,
\be
\begin{split}
\label{eq:finaloffshellOmegaHnochoice}
	S  |_{\Omega\, \mathbb{H}^4} =\, & 16 \pi^2 L^2\, \left(1+ \frac{(b_1 - b_2)^2}{4 b_1 b_2} \right) \sum_{n=0}^\infty \frac{F^{(0,n)}}{L^{2 n}} \\
&+  16 \pi^2 L^2\,  \frac{(b_1 - b_2)^2}{4 b_1 b_2}  \sum_{m=1}^\infty \sum_{n=0}^\infty \frac{F^{(m,n)}}{L^{2 (m+n)}} \frac{(b_1-b_2)^{2 (m-1)}}{ (b_1 + b_2)^{2 (m-1)}}\\
=\, & \frac{4 \pi^2 L^2 (b_1 + b_2)^2}{b_1 b_2} \sum_{m, n=0}^\infty \frac{F^{(m,n)}}{L^{2 (m+n)}} \frac{(b_1-b_2)^{2 m}}{ (b_1 + b_2)^{2 m}} \\
=\, &  \frac{4 \pi^2}{b_1 b_2} F\left((b_1+b_2) L\, X^I; (b_1-b_2)^2, (b_1 + b_2)^2\right) \ .
\end{split}
\ee	
The final answer is in exact agreement with the prediction of \cite{Hristov:2021qsw}, as further discussed in the next section.

\subsection{Introducing $G_N$ and going on-shell}
\label{sec:onshell}
We have so far focused purely on establishing off-shell supersymmetry for the $\Omega\, \mathbb{R}^4$ and $\Omega\, \mathbb{H}^4$  backgrounds, and evaluating the resulting action. The process of going to the Poincar\'e supergravity, which exhibits only the physically relevant symmetries and degrees of freedom, requires gauge fixing all the redundant symmetries in the superconformal group, as well as solving the equations of motion of the auxiliary fields, see \cite{Lauria:2020rhc} and \cite{Bobev:2021oku} for a more extended discussion. In the process of solving the supersymmetry constraints we inadvertently performed some of these steps, but here we complete this procedure in two steps in order to also present a meaningful answer for the on-shell action of $\Omega\, \mathbb{H}^4$.

\subsubsection*{Gauge fixing}
In the process of fixing the supersymmetric configurations, a number of gauge freedoms were already fixed, and we summarize them here for completeness. The special conformal symmetries and the $\SO(1,1)_R$ symmetries were fixed by the choices \eqref{eq:fixK} and \eqref{eq:fixA}, respectively, while the $\SU(2)_R$ symmetry was fixed by the choice \eqref{eq:gaugefix}. These choices, via supersymmetry, lead to fixing the $S$-supersymmetries as in \eqref{eq:s-gauged}, and (as always) the particular choice of backgrounds and the solution of the Killing spinor equations fixed the general coordinate transformations and the $Q$-supersymmetries. Additionally we have the freedom to fix the dilatations, which typically means we are free to set $\chi_\mathrm{H}$ to an arbitrary constant allowing us to introduce a dimensionful constant such as $G_N$ (on top of the FI parameters $g_I$ that are inversely related to the length scale $L$). However, we saw that $\chi_\mathrm{H}$ itself drops out of the action, and therefore we have no interest in its particular value. An alternative (and equally general) approach was outlined in \cite{Hristov:2021qsw}, based on the simple postulate that every order of higher derivatives should come with an additional power of the Newton constant. This point of view was already advocated in \cite{Bobev:2020egg} from an effective theory point of view, since we naturally expect string theory compactifications (or possibly other UV embeddings) to produce higher derivative corrections with a corresponding coupling constant suppression. In turn, it is easy to notice the scalars $X^I_\pm$ appearing in the prepotential can simply be rescaled by
\be
\label{eq:rescaleX}
	X^I_\pm \rightarrow \kappa^{-1}\, X^I_\pm\ , \qquad \kappa^2 := 8 \pi\, G_N\ ,
\ee
which automatically introduces the expected factors of $G_N$ in each individual term in the Lagrangian. 

We are therefore lead to the simple result that, after gauge-fixing, the {\it off-shell} action for $\Omega\, \mathbb{H}^4$ is given by
\be
\label{eq:finaloffshellOmegaH}
	S  |_{\Omega\, \mathbb{H}^4}  =  \frac{4 \pi^2}{b_1 b_2} F\left((b_1+b_2) \tfrac{L}{\kappa}\, X^I; (b_1-b_2)^2, (b_1 + b_2)^2\right) \ .
\ee
Let us stress once again that this result is obtained after the particular choice \eqref{eq:rescaleX}, which even if seemingly logical might not be precisely realized in this way by every possible UV embedding. However, one can always come back to the fully general result in \eqref{eq:finaloffshellOmegaHnochoice} and impose the choice that is relevant in a given situation.

\subsubsection*{Equations of motion}
We are now left with solving all equations of motion in order to find the on-shell action. In reality we have almost completed this task automatically while solving the supersymmetry constraints: (anti-) self-dual two forms drop out of the Einstein equations, and supersymmetry on $\mathbb{R}^4$ and $\mathbb{H}^4$ ensures the Poincar\'e supergravity field equations are automatically solved, see \cite{Hristov:2009uj}. We are then left to check the auxiliary scalar equations of motion, i.e.\ those stemming from the $D$ and $Y^I_{ij}$ fields. 

The $D$ field equation for the particular background value $D=0$, as already mentioned above, serves to relate the scalars $X^I$ to the hyper-K\"ahler potential $\chi_\mathrm{H}$ and can be used to introduce the Newton constant after using the dilatation symmetry gauge fixing to choose $\chi_\mathrm{H}$. We circumvented this procedure here by the rescaling \eqref{eq:rescaleX}, which instead would lead to fixing $\chi_\mathrm{H}$ via the $D$ equation of motion. Since the hypermultiplet sector dropped out of the Lagrangian, we can safely neglect the $D$ equation of motion without any consequences. 

The situation with the $Y^I_{ij}$ equations of motion is different depending on whether we are in the ungauged or in the FI gauged case. In the lack of gauging, $Y^I_{ij}$ is set to zero and a simple check on the equations of motion from the Lagrangian \eqref{eq:Eucl-action} shows that they are automatically satisfied due to the lack of linear terms in $Y^I$. This is enough for us to conclude that $\Omega\, \mathbb{R}^4$ is automatically a supersymmetric {\it solution} of higher derivative gravity. From a practical point of view this was easy to anticipate since the only freedom in the configuration was given by the arbitrary scalars $X^I_\pm$ that are indeed known to remain unfixed in flat space due to the lack of scalar potential. In the presence of FI gauging the exact opposite is true - supersymmetry relates $Y^I$ and $X^I$ as in \eqref{eq:YtoX}, and the resulting equation of motion is non-trivial. In terms of the on-shell degrees of freedom, it means we have to fix the scalars at the minimum of their non-trivial potential, as expected for a (Euclidean) AdS vacuum solution. What supersymmetry has guaranteed for the scalars is that they are constant, and therefore we have evaluated the off-shell action so far without the need to fix the particular values of $X^I$. We are in a position analogous to Sen's entropy function, \cite{Sen:2005wa}, precisely designed to simplify the calculation of the higher derivative action (on a black hole background). Once the coordinate dependence of all fields is fixed on the given configuration, one can proceed to evaluate the action in terms of constants and only at the end extremize with respect to them in order to satisfy all remaining equations of motion. In the present case we only need to extremize the scalars $X^I$ under the gauge fixing constraint \eqref{eq:gaugefix}. We can summarize this discussion by the simple relation between the off-shell and on-shell action,
 \be
	\frac{\partial S^\text{off-shell}}{\partial X^I} |_{\hat{X}^I}  = 0\ , \qquad S^\text{on-shell} = S^\text{off-shell} (\hat{X}^I)\ ,
\ee
under the constraint $g_I X^I = L^{-1}$. We therefore find the formal answer for the on-shell action of $\Omega\, \mathbb{H}^4$,
\be
	S^\text{on-shell}  |_{\Omega\, \mathbb{H}^4}  (b_1, b_2) = \frac{4 \pi^2}{b_1 b_2} F\left((b_1+b_2) \tfrac{L}{\kappa}\, \hat{X}^I; (b_1-b_2)^2, (b_1 + b_2)^2\right) \ ,
\ee
which depends on the particular choice of prepotential. This extremization procedure can be seen as the higher derivative generalization of F-extremization, \cite{Jafferis:2010un,Jafferis:2011zi}, and was discussed at length in \cite{Hristov:2021qsw}. We also give an explicit example of how it works in a holographically motivated setting below, see also \cite{Hristov:2022lcw}.

\section{Towards a proof of \cite{Hristov:2021qsw} and \cite{Hristov:2022lcw}}
\label{sec:on}
Let us now discuss more carefully the relation of the present results with the conjecture of \cite{Hristov:2021qsw}, which we first need to briefly summarize.~\footnote{We are only going to focus on part I of the conjecture in \cite{Hristov:2021qsw} regarding higher derivative corrections. Part II of the conjecture concerns genuine quantum corrections and can be understood from the index theorem as discussed in \cite{Hristov:2021zai}, but further explicit derivation is beyond the present discussion.} The main idea of the conjecture is the prediction of a unifying formula for the off-shell/on-shell action of supersymmetric backgrounds as a sum over contributions from individual fixed points. Assuming a higher derivative action defined by the prepotential \eqref{eq:1}, the action on a supersymmetric background $M_4$ is given by~\footnote{Note that the conjecture was originally presented in Lorentzian conventions, where the prepotential standarly comes with an additional factor of $\i$ that we have stripped off here.}
\bea
\label{eq:conj1}
\begin{split}
	\cF (M_4, \chi^I, \omega)  = &  \sum_{\sigma = 1}^{\chi( M_4)} s_{(\sigma)}\, \cB (\kappa^{-1}\, X^I_{(\sigma)}, \omega_{(\sigma)})\ , \\
	\cB(X^I, \omega) := & \frac{4 \pi^2}{\omega}\, F( X^I; (1-\omega)^2, (1+\omega)^2)\ ,
\end{split}
\eea
where the identification of signs $s_{(\sigma)}$, the geometric deformation parameters $\omega_{(\sigma)}$ and the scalars $X^I_{(\sigma)}$ at each fixed point, together with one overall constraint $\lambda (g_I, \omega, X^I) = 0$, depend on the particular background and go under the called {\it gluing rule}, \cite{Hosseini:2019iad}. In particular, the gluing rule pertaining to the squashed sphere background, called here $\Omega\, \mathbb{H}^4$, known to exhibit a single nut, $\sigma = 1$, was conjectured to be~\footnote{Here we reinsert the length scale $L$, which was set to $1$ in \cite{Hristov:2021qsw}.}
\be
\label{eq:Sgluing}
	M_4 = \Omega\, \mathbb{H}^4\, : \qquad	s_{(1)} = 1\ , \qquad \omega_{(1)} = \omega\ , \qquad X^I_{(1)} = (1 + \omega)\, L\, X^I\ ,
\ee
under the constraint $g_I X^I = L^{-1}$.  Finally, in order to obtain an exact agreement with our present notation and results, we have the relation
\be
\label{eq:newparam}
	\omega = \frac{b_2}{b_1}\ ,
\ee
or the inverse (there is a symmetry under exchange of $b_1$ and $b_2$ and therefore under $\omega \leftrightarrow \omega^{-1}$), in the regular regime $b_2 / b_1 > 0$. It is now immediate to conclude that our result \eqref{eq:finaloffshellOmegaH} for the off-shell action (and therefore also the on-shell action that follows) of $\Omega\, \mathbb{H}^4$ is in agreement with the conjectural formula \eqref{eq:conj1} under the gliung rule \eqref{eq:Sgluing}. As further discussed at length in \cite{Hristov:2021qsw}, the extremization of the action with respect to the deformation $\omega$, also known as squashing parameters, leads to $\hat{\omega} = 1$ such that we recover the empty $\mathbb{H}^4$ vacuum with a round sphere boundary.

We have thus {\it proven} the validity of the proposed gluing rule in \eqref{eq:Sgluing}. It is clearly desirable to prove the general formula \eqref{eq:conj1} at least for a larger set of different supersymmetric backgrounds, and the present results allow us to discuss how such a proof may be devised. Our focus so far have been the two maximally symmetric spaces with a deformation at a single fixed point, and we are now in a position to make some useful conclusions from our explicit calculations. In order to give more precise statements in the particular context,  it is at this point natural to split the discussion of the ungauged and the FI gauged cases.

\subsection{The asymptotically flat case}
\label{subsec:conjflat}
Here we focus on what we have learnt from the $\Omega\, \mathbb{R}^4$ background. At first sight it might seem that we have failed to find anything, given that we could not present a finite action in this example. Indeed the final answer itself in this case is not a subject of prediction from formula \eqref{eq:conj1}, which explicitly requires the existence of a {\it finite} action. We can however make three important remarks about the validity of \eqref{eq:conj1} on asymptotically flat backgrounds in ungauged supergravity, based on the technical results here:
\begin{itemize}
	\item {\it Volume regularization}\\
		In order to obtain a finite action, it is clear that some mechanism of volume regularization is needed. Empty flat space does not provide such a mechanism, but it is known that asymptoptically flat black holes are a class of configurations with finite action. The difference can be physically understood due to the existence of a finite horizon area in the latter examples. This also explains the wide applicability and use of the $\Omega\, \mathbb{R}^4$ in field theory localization on {\it compact} manifolds. 

	\item {\it $\mathbb{W}$-tower contributions}\\
		Even when the volume is regularized in a meaningful way, we observed that all terms carrying higher derivative corrections entirely vanished from the calculation in section \ref{subsec:31}. The reason can be traced back to the nature of the considered deformation, being (anti-) self-dual. Due to the opposite chiralities in the Weyl multiplet definition, \eqref{eq:W-mult}, it becomes a technical requirement that both $T^+$ and $T^-$ are switched on, as otherwise the $\mathbb{W}$-tower of higher derivatives does not contribute. An example of both $T^+$ and $T^-$ switched on are again black hole spacetimes, since Lorentzian signature automatically enforces $T^\pm$ to be related via complex conjugation.

	\item {\it $\mathbb{T}$-tower contributions} \\
		The previous observation is not enough to save the $\mathbb{T}$-tower of higher derivative contributions, which can be seen to vanish as long as we commit ourselves to ungauged supergravity. This already follows from the general requirements in section \ref{sec:susy} regarding the ungauged case (or the $g_I = 0$ limit), and was also observed for black holes in \cite{Butter:2014iwa}.
\end{itemize}
We notice that, apart from the $\Omega\, \mathbb{R}^4$ background, the supersymmetric black holes in Minkowski$_4$ are non-trivial examples that also obey the above observations. In fact single centered black holes in Minkowski are already known to obey \eqref{eq:conj1} with their corresponding gluing rule, see \cite{Hristov:2021qsw}, from a series of previous works, \cite{LopesCardoso:1998tkj,LopesCardoso:1999fsj,LopesCardoso:2000qm}, which served as an inspiration for the general conjecture in first place. It is important here to make the following remark: it was shown in \cite{Hristov:2022pmo} via BPS thermodynamics that asymptotically flat single centered black holes do indeed correspond to $\omega = -1$ (in the conventions of \eqref{eq:conj1}), such that the potential $\mathbb{T}$-tower of corrections vanishes. However, it would be misleading to expect that potentially new asymptotically flat examples with $\omega \neq -1$ might lead to non-vanishing $\mathbb{T}$-corrections, as this is not possible. It is then a question of further study to understand if this automatically implies that $\omega = -1$ for {\it all} backgrounds in ungauged supergravity, or one should rather specify a refined version of \eqref{eq:conj1}.

\subsection{Nut action in Euclidean AdS}
Here we consider general supersymmetric asymptotically locally Euclidean AdS (or $\mathbb{H}^4$) backgrounds, and discuss how \eqref{eq:conj1} can be proven with the help of our explicit calculations on $\Omega\, \mathbb{H}^4$, leaving the actual accomplishment of such a proof as a future goal. For this purpose we need to come back to the Gibbons-Hawking picture of nuts and bolts, \cite{Gibbons:1979xm}, and its application to supersymmetric AdS backgrounds developed in \cite{Martelli:2012sz,Farquet:2014kma,BenettiGenolini:2016tsn,BenettiGenolini:2019jdz}. Using the fixed point set of the canonical isometry $\xi$ (produced from the Killing spinor bilinears), the restriction of \eqref{eq:conj1} to the two derivative theory with no additional matter was shown in \cite{BenettiGenolini:2019jdz} to apply for all supersymmetric asymptotically AdS backgrounds in the minimal Poincar\'e supergravity with cosmological constant.~\footnote{See also \cite{Genolini:2021urf} for the analogous proof in {\it on-shell} four derivative minimal supergravity.} The conceptual idea of this proof, which we sketch next, is also directly applicable to the generalization with extra matter and higher derivatives. We can divide the analysis in three main steps:
\begin{enumerate}
	\item {\it Reduction on a base}\\
		Using the isometry $\xi$, any background space $M_4$ can be written in local coordinates as a circle fibration over a base $B_3$. The full action can then be reduced on the isometry down to a three-dimensional action on the base, and using the equations of motion the resulting {\it on-shell} action is shown to be exact. Stokes' theorem leads to an integral over the boundary of $B_3$, which consists of the conformal boundary at inifinity and the boundaries of the neighborhoods around the fixed point set (nuts and bolts) of $\xi$.

	\item {\it Conformal boundary contributions}\\
		The evaluation of the conformal boundary contributions requires holographic renormalization, which in two derivative minimal supergravity only requires the standard Gibbons-Hawking-York counterterm, see \eqref{eq:Hct}. Due to an additional freedom in the term known as the {\it nut potential}, see \cite{BenettiGenolini:2019jdz}, the relative contributions between the conformal boundary and the fixed points can be freely adjusted, allowing for the natural choice of putting all asymptotic contributions to zero and leaving the full answer to be determined on the nuts and bolts.

	\item {\it Fixed point contributions}\\
		The final answer is thus entirely derived from the integral on (the boundary around) the fixed point set, ultimately leading to a sum over nuts and bolts. The sum over nuts is precisely the restriction of \eqref{eq:conj1} to the absence of matter and higher derivative corrections. 
\end{enumerate}

The same three steps can be undertaken in the general case we consider here, with some important technical differences that we emphasize.  Repeating the first step requires starting from the fully {\it off-shell} action we work with, \eqref{eq:Eucl-action}, performing the reduction along the isometry $\xi$ and rearrangement of the resulting terms without the use of the equations of motion. The second step is technically independent of the first one and requires the systematic application of holographic renormalization on theories with additional matter and higher derivative couplings. We can roughly differentiate the terms in the action in two main groups: curvature invariants and field strength/two form terms, both of these coupled with (composite) scalars. In the absence of scalar coupling, the curvature invariants generically require respective counterterms, see \cite{Bobev:2021oku}, while the field strength/two form terms are already vanishing at the boundary. The addition of potentially non-constant scalars requires further understanding and leaves open the question of potential finite counterterms that needs to be addressed.~\footnote{Due to the fact that the scalars were shown to be constant on $\Omega\, \mathbb{H}^4$ we could determine the relevant counterterms and evaluate the action without such issues, but this simplification cannot hold for all supersymmetric backgrounds.} Finally, the evaluation of the nut/bolt contributions should follow straightforwardly given the completion of the first step, since there do not appear to be additional complications. 

Resolving these points rigorously is left for future studies. However, what gives us strong confidence in the expected outcome of this computation is the successful evaluation of the off-shell action on $\Omega\, \mathbb{H}^4$ and its precise agreement with \eqref{eq:conj1}. The choice of particular background in this case allowed us to circumvent many of the subtleties mentioned above, while still giving us a clean proof that a single nut contribution can be understood by the above logic. It would also be interesting to repeat this on a background exhibiting a single bolt, as the ones discussed in \cite{Toldo:2017qsh}, in order to complete the general expression \eqref{eq:conj1}.

\subsection{11d on S$^7 / \mathbb{Z}_k$ from holography}
Finally, let us also present an explicit example where the two towers of inifinite derivative corrections can be seen to play a role. Based on the conjectural expression for the off-shell action of $\Omega\, \mathbb{H}^4$, which we have now proven, \cite{Hristov:2022lcw} applied what was described as {\it holographic bootstrap} in order to find the higher derivative prepotential resulting from 11d supergravity reduction on  S$^7 / \mathbb{Z}_k$. This simply amounts to imposing the holographic match with the squashed sphere partition function of ABJM theory, \cite{Aharony:2008ug}, which was evaluated via supersymmetric localization beyond the large $N$ limit in \cite{Fuji:2011km,Marino:2011eh,Nosaka:2015iiw,Hatsuda:2016uqa,Chester:2021gdw}. Referring to \cite{Hristov:2022lcw} for more details, this holographic comparison ultimately leads to the following prediction for the form of the higher derivative prepotential \eqref{eq:1} and FI gauging,~\footnote{We are again converting to Euclidean notation and reinserting the length scale $L$ where appropriate.}
\be
\label{eq:S7prepot}
	F = 2 \sqrt{X^0 X^1 X^2 X^3}\, \, \sum_{n=0}^\infty f_n \left( \frac{ k_{\mathbb{W}} (X) A_{\mathbb{W}} + k_{\mathbb{T}} (X) A_{\mathbb{T}}}{64\, X^0 X^1 X^2 X^3} \right)^n\ , \qquad g_I = L^{-1}\ , \forall I\ ,
\ee
where
\be
\label{eq:k}
\begin{split}
k_\mathbb{W} (X) & := -2 \sum_{I<J} X^I X^J\ , \\
k_\mathbb{T} (X)  := \sum_I  (X^I)^2 & - \frac{(X^0+X^1-X^2 - X^3) (X^0-X^1+X^2-X^3) (X^0- X^1 - X^2+X^3)}{\sum_I X^I}\ ,
\end{split}
\ee
and coefficients $f_n$ given in terms of the numbers $N$ and $k$ as in \cite{Hristov:2022lcw}, related to the 11d solution AdS$_4 \times$S$^7/\mathbb{Z}_k$ representing the near-horizon limit of $N$ coincident M2-branes,
\bea
\label{eq:identifyNk}
\begin{split}
	\frac{L^2\, f_0}{4 G_N} &= \frac{\sqrt{2 k}}{3}\, \left( N - \frac{k}{24} \right)^{3/2} , \\
	2 \pi f_1 &= \left(-\frac1{2 k} \right)\,  \frac{\sqrt{2 k}}{3}\, \left( N - \frac{k}{24} \right)^{1/2} , \\
& ...  \\
\frac{2 \pi L^{2 (1-n)}\, f_n}{\kappa^{2 (1-n)}} &= \left( \frac{(2 n-5)!!\, 3 }{n!\, (6 k)^n}   \right)\, \frac{\sqrt{2 k}}{3}\, \left( N - \frac{k}{24} \right)^{3/2-n} .
\end{split}
\eea
The prepotential can thus be recognized as the Taylor expansion of the simple power function
\be
	F =\frac{\kappa^2\, \sqrt{2 k X^0 X^1 X^2 X^3}}{3 \pi L^2}\, \,  \left(N - \frac{k}{24} -\frac{L^2}{3 k}\, \frac{  k_{\mathbb{W}} (X) A_{\mathbb{W}} + k_{\mathbb{T}} (X) A_{\mathbb{T}}}{64 \kappa^2\, X^0 X^1 X^2 X^3} \right)^{3/2}\ .
\ee

The off-shell action of $\Omega\, \mathbb{H}^4$, using the parametrization \eqref{eq:newparam}, is then evaluated to
\be
\label{eq:abjmsphere}
	S  |_{\Omega\, \mathbb{H}^4}  = \frac{4 \pi (1+\omega)^2}{3\, \omega} \sqrt{2 k X^0 X^1 X^2 X^3}\left(N - \frac{k}{24} - \frac{ k_{\mathbb{W}} (X) (1-\omega)^2 + k_{\mathbb{T}} (X) (1+\omega)^2 }{192 k (1+\omega)^2\, X^0 X^1 X^2 X^3} \right)^{3/2}\ ,
\ee
under the constraint $\sum_I X^I = 1$. Upon writing this in the natural field theory variables, $X^I = \tfrac12 \Delta_i$,~\footnote{This map, which here is claimed to hold {\it off-shell} with no derivation, is standardly established in holography {\it on-shell} using an explicit solution with position dependent scalars as in the two derivative example of \cite{Freedman:2013oja}.} the result can be recognized as the leading term in an Airy function expansion, see \cite{Hristov:2022lcw}. It is precisely this relation that lead to the identification of \eqref{eq:S7prepot}- \eqref{eq:identifyNk}. Note that in this case it is easy to perform the extremization of the off-shell action explicitly, recovering $\hat{X}^I = 1/4, \forall I$ that already holds at leading order. The on-shell action is thus given by
 \be
\label{eq:abjmsphereonshell}
	S  |_{\Omega\, \mathbb{H}^4} (\hat{X})  = \frac{ \sqrt{2 k} \pi (1+\omega)^2}{12\, \omega}\left(N - \frac{k}{24} + \frac{ 3 (1-\omega)^2 - (1+\omega)^2 }{3 k (1+\omega)^2} \right)^{3/2}\ .
\ee

Having now proven the direct relation between \eqref{eq:S7prepot} and \eqref{eq:abjmsphere}, the results reported in \cite{Hristov:2022lcw} concerning the sphere partition function of ABJM and the holographic prediction of the higher derivative prepotential are on a more solid ground. However, the generic form of other partition functions given in \cite{Hristov:2022lcw} still relies on the full validity of \eqref{eq:conj1} combined with the gluing rules for the particular background in question. It is a non-trivial holographic check that both the twisted and the identity gluing rules (see \cite{Hosseini:2019iad,Hristov:2021qsw}) lead to the correct {\it finite} $N$ answer for the topologically twisted and superconformal indices of ABJM, respectively, as numerically verified in \cite{Bobev:2022jte} and \cite{Bobev:2022wem}. Additionally, at {\it large} $N$, the refined topologically twisted index of ABJM was also shown in \cite{Hosseini:2022vho} to agree with the proposed gluing rules applied to \eqref{eq:conj1}, generalizing the analytic result of \cite{Benini:2015eyy}.

\section{Outlook and extensions}
\label{sec:out}
There are a number of open questions related to the conjecture of \cite{Hristov:2021qsw} and higher derivative supergravity left for future studies. In the previous section we have discussed at some length the evaluation of the action for supersymmetric backgrounds exhibiting nuts, or fixed points, and the remaining steps to complete the proof of \eqref{eq:conj1}. A natural extension would be to also include possible bolts, such that \eqref{eq:conj1} is completed. In turn one can also investigate the quantum corrections coming on top of the higher derivative terms, see again \cite{Hristov:2021zai,Hristov:2021qsw}, such that we complete the {\it quantum} gravitational block (conjectured in \cite{Hristov:2022lcw} via holography to be the complete Airy function for the AdS$_4 \times$S$^7/\mathbb{Z}_k$ example). It would likewise be very interesting to fully understand the analogous gravitational building blocks in different dimensions (see \cite{Hosseini:2019iad,Faedo:2021nub}) and even directly inside string compactifications via the internal geometry (see \cite{Boido:2022mbe}). We hope to come back to all of these problems. The present work does allow a more detailed understanding of two particular extensions, while remaining in 4d higher derivative gauged supergravity, to which we turn next.

\subsection{Adding hypermultiplets}
\label{sec:hyp}

The present results for the $\Omega\, \mathbb{R}^4$ and $\Omega\, \mathbb{H}^4$ backgrounds were derived using the choice of a single compensating hypermultiplet, which is invisible in the on-shell formulation. However, much of our analysis is directly applicable to the case of arbitrary hypers, which means that we can now conceptually understand the impact of on-shell hypermultiplets to the calculation. As already mentioned in section \ref{sec:susy}, maximal supersymmetry in the superconformal hypermultiplet sector leads to the conditions \eqref{eq:genhyp}, which can always be solved on a given manifold. At this stage the discussion can be split into two parts that can be considered independently.

First, we should stress that the hypermultiplet geometry is now encoded in the hyper-K\"ahler potential $\chi_\mathrm{H}$, as discussed in \cite{deWit:1999fp,deWit:2001bk}. This function plays an analogous role to the prepotential \eqref{eq:1} that specifies the vector multiplet geometry. Importantly, $\chi_\mathrm{H}$ can also receive corrections from string compactifications, as discussed in e.g.\ \cite{Robles-Llana:2006vct}, and these corrections are completely independent from the higher derivative invariants $\mathbb{W, T}$ that we have focused on. Explicit correction to the universal hypermultiplet for example were worked out in \cite{Antoniadis:2003sw,Anguelova:2004sj}, but more general results have not appeared in the literature.

Second, due to the independence of the hypermultiplet sector, the only way our results are going to be changed is via the presence of a gauged isometry on the hypermultiplet moduli space. As expalined above, it is hard to work with an explicit example in the superconformal formalism, but it is relatively easy to see the consequence of gauging if we look at the Poincar\'e theory (see \cite{deWit:1999fp,deWit:2001bk} for the relation of the two formalisms in the hypermultiplet sector). Since we imposed maximal supersymmetry in this sector for both $\Omega\, \mathbb{R}^4$ and $\Omega\, \mathbb{H}^4$, we can directly borrow the results of \cite{Hristov:2009uj} and \cite{Hristov:2010eu} in this case. The gauging of a given hypermultiplet isometry is parametrized by a killing vector $k^u_I (q)$ (we remind the reader that the hypermultiplet scalars are denoted by $q^u$), selecting the particular linear combination of gauge fields $W^I_\mu$ used for the gauging. The resulting moment maps $P_I (q)$ generalize the FI parameters $g_I$, and the vevs $\hat{q}^u$ at the minimum of the resulting scalar potential selects whether one can realize the $\mathbb{R}^4$ ($P_I (\hat{q}) = 0$) or the $\mathbb{H}^4$ ($P_I (\hat{q}) \neq 0$) vacuum. The only additional difference in the supersymmetry backgrounds is the additional constraint
\be
\label{eq:susyHiggs}
	k^u_I (\hat{q}) X^I = 0\ ,
\ee  
which needs to hold at the vacuum. As further observed in \cite{Hosseini:2017fjo,Hosseini:2020vgl}, this constraint is satisfied automatically for compact isometries since $k^u_I (\hat{q}) = 0$, but leads to an additional non-trivial constraint on the vector multiplet scalars for non-compact isometries where $k^u_I (\hat{q}) \neq 0$. The additional constraint \eqref{eq:susyHiggs} can be seen as a direct consequence of a supersymmetry-preserving Higgs mechanism that takes place on the background configuration, see \cite{Hristov:2010eu}. Therefore, our expectation is that the final answer for the supersymmetric configurations and the resulting evaluation of the action remains the same, upon the replacement of $g_I$ with $P_I (\hat{q})$ and the additional constraint \eqref{eq:susyHiggs}.

We should however stress that we are not aware of explicit supergravity models constructed from string theory flux compactification that include higher derivative/quantum corrections and non-trivial gauging, such that the above discussion is somewhat abstract at the moment. It is worth pointing out that a {\it holographic bootstrap} of the Lagrangian, similar to the one employed to determine \eqref{eq:S7prepot} above, is unfortunately not applicable for the hypermultiplet sector since it is not readily visible in the final off/on-shell action. Ultimately the difference is due to the fact that keeping track of flavor symmetries in field theory allows us a direct relation to background vector multiplets in supergravity, while similar relation does not exist for the hypermultiplet sector.

\subsection{Conical defects}
The interest in allowing conical defects to asymptotically AdS solutions in various dimensions has recently spiked due to the original work of \cite{Ferrero:2020laf,Ferrero:2020twa} on the so-called spindle geometry, which was also shown to make sense holographically. The gluing rules applicable to black holes with spindle horizons were later proposed in \cite{Hosseini:2021fge} and generalized in various directions in \cite{Faedo:2021nub} and \cite{Boido:2022mbe}. From the point of view of the present work, focusing on the 4d perspective, the black spindles are backgrounds exhibiting two different fixed points, near which the metric is locally $\mathbb{C}\times \mathbb{C}/\mathbb{Z}_{n_1}$ and $\mathbb{C}\times \mathbb{C}/\mathbb{Z}_{n_2}$, respectively. The conical deficit angles $n_1$ and $n_2$ then play an important role in characterizing the geometry and the full solution, see again \cite{Ferrero:2020laf,Ferrero:2020twa} and references thereof.

In the context of the $\Omega\, \mathbb{H}^4$ background that we focused on, adding conical deficit angles is practically a rather simple exercise, which we can perform in the following way. First, notice that the $\mathbb{H}^4$ metric,
\be
	{\rm d} s^2 = \frac{{\rm d} r^2}{1+\tfrac{r^2}{L^2}} + r^2 \left( {\rm d} \theta^2 + \cos^2 \theta {\rm d} \varphi_1^2 + \sin^2 \theta {\rm d} \varphi_2^2 \right) \ ,	
\ee
near the fixed point at the center ($r = 0$) exhibits the $\mathbb{C}\times \mathbb{C}$ metric in two copies of polar coordinates, c.f.\ \eqref{eq:twopolars}. The angles $\varphi_1$ and $\varphi_2$ are  precisely the angles on the two complex planes, and therefore we can simply introduce two {\it independent} conical deficit angles by changing their periodicities,
\be
\label{eq:conical}
	\varphi_1 \in [0, \frac{2 \pi}{p} )\ , \qquad \varphi_2 \in [0, \frac{2 \pi}{q} )\ .
\ee
Importantly, both $\varphi_1$ and $\varphi_2$ are isometries not only of the metric, but the complete set of background fields \eqref{eq:fullOH}, and furthermore the explicit Killing spinors are independent of $\varphi_1, \varphi_2$ as shown in \cite{Martelli:2011fu,Farquet:2014kma} (see section \ref{subsec:omegas}). We are therefore allowed to take arbitrary values for $p$ and $q$ in \eqref{eq:conical} without changing anything else in the background \eqref{eq:fullOH} and without breaking any of the original symmetries and supersymmetries. The resulting spacetime, which corresponds to $\Omega\, \mathbb{H}^4/ (\mathbb{Z}_p \times \mathbb{Z}_q)$ then exhibits a single nut that is locally $\mathbb{C}/\mathbb{Z}_p \times \mathbb{C}/\mathbb{Z}_q$. The evaluation of the off-shell action is also in complete analogy, with the more general periodicities simply appearing in the final integration,
\be
\label{eq:finaloffshellOmegaHconical}
	S  |_{\Omega\, \mathbb{H}^4/ (\mathbb{Z}_p \times \mathbb{Z}_q)}  = \frac{4 \pi^2}{p\, q}  \frac{1}{b_1 b_2} F\left((b_1+b_2) \tfrac{L}{\kappa}\, X^I; (b_1-b_2)^2, (b_1 + b_2)^2\right) \ ,
\ee
and the on-shell action following upon extremization, as before. It is therefore natural to expect that the conical deficit angles $p$ and $q$ at a given nut to appear in the denominator of the building block, 
\be
	\cB_{p,q} (X^I, \omega) := \frac{4 \pi^2}{p\, q\, \omega}\, F( X^I; (1-\omega)^2, (1+\omega)^2)\ ,
\ee
which enters in the factorization formula \eqref{eq:conj1}, where each different fixed point $\sigma$ can a priori come with independent integers $p_{(\sigma)}$ and $q_{(\sigma)}$. It would be interesting to explore in more detail the higher derivative corrections on the black spindles and derive rigorously the corresponding gluing rules. We leave this for future study.

\section*{Acknowledgements}
I am very grateful to Nikolay Bobev, Anthony Charles, Fridrik Freyr Gautason, Seyed Morteza Hosseini, Stefanos Katmadas, Ivano Lodato, Hugo Looyestijn, Achilleas Passias, Valentin Reys, Chiara Toldo, Alessandro Tomasiello, Stefan Vandoren, and Alberto Zaffaroni for collaboration and discussions on related subjects. I am supported in part by the Bulgarian NSF grants N28/5 and KP-06-N 38/11.

\bibliographystyle{JHEP}
\bibliography{omega.bib}

\providecommand{\href}[2]{#2}\begingroup\raggedright\begin{thebibliography}{10}

\bibitem{Hristov:2021qsw}
K.~Hristov, {\it {4d $ \mathcal{N} $ = 2 supergravity observables from
  Nekrasov-like partition functions}},  {\em JHEP} {\bf 02} (2022) 079,
  [\href{http://arxiv.org/abs/2111.06903}{{\tt arXiv:2111.06903}}].

\bibitem{Hristov:2022lcw}
K.~Hristov, {\it {ABJM at finite N via 4d supergravity}},  {\em JHEP} {\bf 10}
  (2022) 190, [\href{http://arxiv.org/abs/2204.02992}{{\tt arXiv:2204.02992}}].

\bibitem{Gibbons:1976ue}
G.~Gibbons and S.~Hawking, {\it {Action Integrals and Partition Functions in
  Quantum Gravity}},  {\em Phys. Rev. D} {\bf 15} (1977) 2752--2756.

\bibitem{Gibbons:1979xm}
G.~W. Gibbons and S.~W. Hawking, {\it {Classification of Gravitational
  Instanton Symmetries}},  {\em Commun. Math. Phys.} {\bf 66} (1979) 291--310.

\bibitem{Chamblin:1998pz}
A.~Chamblin, R.~Emparan, C.~V. Johnson, and R.~C. Myers, {\it {Large N phases,
  gravitational instantons and the nuts and bolts of AdS holography}},  {\em
  Phys. Rev. D} {\bf 59} (1999) 064010,
  [\href{http://arxiv.org/abs/hep-th/9808177}{{\tt hep-th/9808177}}].

\bibitem{Martelli:2012sz}
D.~Martelli, A.~Passias, and J.~Sparks, {\it {The supersymmetric NUTs and bolts
  of holography}},  {\em Nucl. Phys. B} {\bf 876} (2013) 810--870,
  [\href{http://arxiv.org/abs/1212.4618}{{\tt arXiv:1212.4618}}].

\bibitem{Farquet:2014kma}
D.~Farquet, J.~Lorenzen, D.~Martelli, and J.~Sparks, {\it {Gravity duals of
  supersymmetric gauge theories on three-manifolds}},  {\em JHEP} {\bf 08}
  (2016) 080, [\href{http://arxiv.org/abs/1404.0268}{{\tt arXiv:1404.0268}}].

\bibitem{BenettiGenolini:2016tsn}
P.~Benetti~Genolini, D.~Cassani, D.~Martelli, and J.~Sparks, {\it {Holographic
  renormalization and supersymmetry}},  {\em JHEP} {\bf 02} (2017) 132,
  [\href{http://arxiv.org/abs/1612.06761}{{\tt arXiv:1612.06761}}].

\bibitem{Toldo:2017qsh}
C.~Toldo and B.~Willett, {\it {Partition functions on 3d circle bundles and
  their gravity duals}},  {\em JHEP} {\bf 05} (2018) 116,
  [\href{http://arxiv.org/abs/1712.08861}{{\tt arXiv:1712.08861}}].

\bibitem{BenettiGenolini:2019jdz}
P.~Benetti~Genolini, J.~M. Perez Ipi\~na, and J.~Sparks, {\it {Localization of
  the action in AdS/CFT}},  {\em JHEP} {\bf 10} (2019) 252,
  [\href{http://arxiv.org/abs/1906.11249}{{\tt arXiv:1906.11249}}].

\bibitem{Belavin:1984vu}
A.~A. Belavin, A.~M. Polyakov, and A.~B. Zamolodchikov, {\it {Infinite
  Conformal Symmetry in Two-Dimensional Quantum Field Theory}},  {\em Nucl.
  Phys. B} {\bf 241} (1984) 333--380.

\bibitem{Nekrasov:2002qd}
N.~A. Nekrasov, {\it {Seiberg-Witten prepotential from instanton counting}},
  {\em Adv. Theor. Math. Phys.} {\bf 7} (2003), no.~5 831--864,
  [\href{http://arxiv.org/abs/hep-th/0206161}{{\tt hep-th/0206161}}].

\bibitem{Nekrasov:2003vi}
N.~A. Nekrasov, {\it {Localizing gauge theories}},  in {\em {14th International
  Congress on Mathematical Physics}}, 7, 2003.

\bibitem{Beem:2012mb}
C.~Beem, T.~Dimofte, and S.~Pasquetti, {\it {Holomorphic Blocks in Three
  Dimensions}},  {\em JHEP} {\bf 12} (2014) 177,
  [\href{http://arxiv.org/abs/1211.1986}{{\tt arXiv:1211.1986}}].

\bibitem{Festuccia:2018rew}
G.~Festuccia, J.~Qiu, J.~Winding, and M.~Zabzine, {\it {Twisting with a Flip
  (the Art of Pestunization)}},  {\em Commun. Math. Phys.} {\bf 377} (2020),
  no.~1 341--385, [\href{http://arxiv.org/abs/1812.06473}{{\tt
  arXiv:1812.06473}}].

\bibitem{Pestun:2007rz}
V.~Pestun, {\it {Localization of gauge theory on a four-sphere and
  supersymmetric Wilson loops}},  {\em Commun. Math. Phys.} {\bf 313} (2012)
  71--129, [\href{http://arxiv.org/abs/0712.2824}{{\tt arXiv:0712.2824}}].

\bibitem{Pestun:2016zxk}
V.~Pestun et~al., {\it {Localization techniques in quantum field theories}},
  {\em J. Phys. A} {\bf 50} (2017), no.~44 440301,
  [\href{http://arxiv.org/abs/1608.02952}{{\tt arXiv:1608.02952}}].

\bibitem{Hosseini:2019iad}
S.~M. Hosseini, K.~Hristov, and A.~Zaffaroni, {\it {Gluing gravitational blocks
  for AdS black holes}},  {\em JHEP} {\bf 12} (2019) 168,
  [\href{http://arxiv.org/abs/1909.10550}{{\tt arXiv:1909.10550}}].

\bibitem{Nekrasov:2003rj}
N.~Nekrasov and A.~Okounkov, {\it {Seiberg-Witten theory and random
  partitions}},  {\em Prog. Math.} {\bf 244} (2006) 525--596,
  [\href{http://arxiv.org/abs/hep-th/0306238}{{\tt hep-th/0306238}}].

\bibitem{Plebanski:1976gy}
J.~F. Plebanski and M.~Demianski, {\it {Rotating, charged, and uniformly
  accelerating mass in general relativity}},  {\em Annals Phys.} {\bf 98}
  (1976) 98--127.

\bibitem{Alonso-Alberca:2000zeh}
N.~Alonso-Alberca, P.~Meessen, and T.~Ortin, {\it {Supersymmetry of topological
  Kerr-Newman-Taub-NUT-AdS space-times}},  {\em Class. Quant. Grav.} {\bf 17}
  (2000) 2783--2798, [\href{http://arxiv.org/abs/hep-th/0003071}{{\tt
  hep-th/0003071}}].

\bibitem{Martelli:2011fu}
D.~Martelli, A.~Passias, and J.~Sparks, {\it {The gravity dual of
  supersymmetric gauge theories on a squashed three-sphere}},  {\em Nucl. Phys.
  B} {\bf 864} (2012) 840--868, [\href{http://arxiv.org/abs/1110.6400}{{\tt
  arXiv:1110.6400}}].

\bibitem{Klare:2013dka}
C.~Klare and A.~Zaffaroni, {\it {Extended Supersymmetry on Curved Spaces}},
  {\em JHEP} {\bf 10} (2013) 218, [\href{http://arxiv.org/abs/1308.1102}{{\tt
  arXiv:1308.1102}}].

\bibitem{Lauria:2020rhc}
E.~Lauria and A.~Van~Proeyen, {\em {${\cal N}=2$ Supergravity in $D=4,5,6$
  Dimensions}}, vol.~966.
\newblock Springer, 2020.

\bibitem{deWit:2017cle}
B.~de~Wit and V.~Reys, {\it {Euclidean supergravity}},  {\em JHEP} {\bf 12}
  (2017) 011, [\href{http://arxiv.org/abs/1706.04973}{{\tt arXiv:1706.04973}}].

\bibitem{Bergshoeff:1980is}
E.~Bergshoeff, M.~de~Roo, and B.~de~Wit, {\it {Extended Conformal
  Supergravity}},  {\em Nucl. Phys. B} {\bf 182} (1981) 173--204.

\bibitem{Butter:2013lta}
D.~Butter, B.~de~Wit, S.~M. Kuzenko, and I.~Lodato, {\it {New higher-derivative
  invariants in N=2 supergravity and the Gauss-Bonnet term}},  {\em JHEP} {\bf
  12} (2013) 062, [\href{http://arxiv.org/abs/1307.6546}{{\tt
  arXiv:1307.6546}}].

\bibitem{Hristov:2009uj}
K.~Hristov, H.~Looyestijn, and S.~Vandoren, {\it {Maximally supersymmetric
  solutions of D=4 N=2 gauged supergravity}},  {\em JHEP} {\bf 11} (2009) 115,
  [\href{http://arxiv.org/abs/0909.1743}{{\tt arXiv:0909.1743}}].

\bibitem{Hama:2012bg}
N.~Hama and K.~Hosomichi, {\it {Seiberg-Witten Theories on Ellipsoids}},  {\em
  JHEP} {\bf 09} (2012) 033, [\href{http://arxiv.org/abs/1206.6359}{{\tt
  arXiv:1206.6359}}]. [Addendum: JHEP 10, 051 (2012)].

\bibitem{Bobev:2019ylk}
N.~Bobev, F.~F. Gautason, and K.~Hristov, {\it {Holographic dual of the
  $\Omega$ -background}},  {\em Phys. Rev. D} {\bf 100} (2019), no.~2 021901,
  [\href{http://arxiv.org/abs/1903.05095}{{\tt arXiv:1903.05095}}].

\bibitem{Skenderis:2002wp}
K.~Skenderis, {\it {Lecture notes on holographic renormalization}},  {\em
  Class. Quant. Grav.} {\bf 19} (2002) 5849--5876,
  [\href{http://arxiv.org/abs/hep-th/0209067}{{\tt hep-th/0209067}}].

\bibitem{Bobev:2020egg}
N.~Bobev, A.~M. Charles, K.~Hristov, and V.~Reys, {\it {The Unreasonable
  Effectiveness of Higher-Derivative Supergravity in AdS$_4$ Holography}},
  {\em Phys. Rev. Lett.} {\bf 125} (2020), no.~13 131601,
  [\href{http://arxiv.org/abs/2006.09390}{{\tt arXiv:2006.09390}}].

\bibitem{Bobev:2021oku}
N.~Bobev, A.~M. Charles, K.~Hristov, and V.~Reys, {\it {Higher-Derivative
  Supergravity, AdS$_4$ Holography, and Black Holes}},
  \href{http://arxiv.org/abs/2106.04581}{{\tt arXiv:2106.04581}}.

\bibitem{Genolini:2021urf}
P.~B. Genolini and P.~Richmond, {\it {The Supersymmetry of Higher-Derivative
  Supergravity in AdS$_4$ Holography}},
  \href{http://arxiv.org/abs/2107.04590}{{\tt arXiv:2107.04590}}.

\bibitem{Jafferis:2010un}
D.~L. Jafferis, {\it {The Exact Superconformal R-Symmetry Extremizes Z}},  {\em
  JHEP} {\bf 05} (2012) 159, [\href{http://arxiv.org/abs/1012.3210}{{\tt
  arXiv:1012.3210}}].

\bibitem{Jafferis:2011zi}
D.~L. Jafferis, I.~R. Klebanov, S.~S. Pufu, and B.~R. Safdi, {\it {Towards the
  F-Theorem: N=2 Field Theories on the Three-Sphere}},  {\em JHEP} {\bf 06}
  (2011) 102, [\href{http://arxiv.org/abs/1103.1181}{{\tt arXiv:1103.1181}}].

\bibitem{Freedman:2013oja}
D.~Z. Freedman and S.~S. Pufu, {\it {The holography of $F$-maximization}},
  {\em JHEP} {\bf 03} (2014) 135, [\href{http://arxiv.org/abs/1302.7310}{{\tt
  arXiv:1302.7310}}].

\bibitem{Binder:2021euo}
D.~J. Binder, D.~Z. Freedman, S.~S. Pufu, and B.~Zan, {\it {The Holographic
  Contributions to the Sphere Free Energy}},
  \href{http://arxiv.org/abs/2107.12382}{{\tt arXiv:2107.12382}}.

\bibitem{Zan:2021ftf}
B.~Zan, D.~Z. Freedman, and S.~S. Pufu, {\it {The $ \mathcal{N} $ = 2
  prepotential and the sphere free energy}},  {\em JHEP} {\bf 06} (2022) 045,
  [\href{http://arxiv.org/abs/2112.06931}{{\tt arXiv:2112.06931}}].

\bibitem{deWit:2011gk}
B.~de~Wit and M.~van Zalk, {\it {Electric and magnetic charges in N=2 conformal
  supergravity theories}},  {\em JHEP} {\bf 10} (2011) 050,
  [\href{http://arxiv.org/abs/1107.3305}{{\tt arXiv:1107.3305}}].

\bibitem{Hristov:2016vbm}
K.~Hristov, S.~Katmadas, and I.~Lodato, {\it {Higher derivative corrections to
  BPS black hole attractors in 4d gauged supergravity}},  {\em JHEP} {\bf 05}
  (2016) 173, [\href{http://arxiv.org/abs/1603.00039}{{\tt arXiv:1603.00039}}].

\bibitem{deWit:1999fp}
B.~de~Wit, B.~Kleijn, and S.~Vandoren, {\it {Superconformal hypermultiplets}},
  {\em Nucl. Phys. B} {\bf 568} (2000) 475--502,
  [\href{http://arxiv.org/abs/hep-th/9909228}{{\tt hep-th/9909228}}].

\bibitem{deWit:2001bk}
B.~de~Wit, M.~Rocek, and S.~Vandoren, {\it {Gauging isometries on hyperKahler
  cones and quaternion Kahler manifolds}},  {\em Phys. Lett. B} {\bf 511}
  (2001) 302--310, [\href{http://arxiv.org/abs/hep-th/0104215}{{\tt
  hep-th/0104215}}].

\bibitem{Bobev:2020pjk}
N.~Bobev, A.~M. Charles, and V.~S. Min, {\it {Euclidean Black Saddles and AdS4
  Black Holes}},  \href{http://arxiv.org/abs/2006.01148}{{\tt
  arXiv:2006.01148}}.

\bibitem{Butter:2014iwa}
D.~Butter, B.~de~Wit, and I.~Lodato, {\it {Non-renormalization theorems and N=2
  supersymmetric backgrounds}},  {\em JHEP} {\bf 03} (2014) 131,
  [\href{http://arxiv.org/abs/1401.6591}{{\tt arXiv:1401.6591}}].

\bibitem{Banerjee:2016qvj}
N.~Banerjee, S.~Bansal, and I.~Lodato, {\it {The Resolution of an Entropy
  Puzzle for 4D non-BPS Black Holes}},  {\em JHEP} {\bf 05} (2016) 142,
  [\href{http://arxiv.org/abs/1602.05326}{{\tt arXiv:1602.05326}}].

\bibitem{York:1986it}
J.~W. York, Jr., {\it {Black hole thermodynamics and the Euclidean Einstein
  action}},  {\em Phys. Rev. D} {\bf 33} (1986) 2092--2099.

\bibitem{Braden:1990hw}
H.~W. Braden, J.~D. Brown, B.~F. Whiting, and J.~W. York, Jr., {\it {Charged
  black hole in a grand canonical ensemble}},  {\em Phys. Rev. D} {\bf 42}
  (1990) 3376--3385.

\bibitem{Sen:2005wa}
A.~Sen, {\it {Black hole entropy function and the attractor mechanism in higher
  derivative gravity}},  {\em JHEP} {\bf 09} (2005) 038,
  [\href{http://arxiv.org/abs/hep-th/0506177}{{\tt hep-th/0506177}}].

\bibitem{Hristov:2021zai}
K.~Hristov and V.~Reys, {\it {Factorization of log-corrections in
  AdS$_4$/CFT$_3$ from supergravity localization}},
  \href{http://arxiv.org/abs/2107.12398}{{\tt arXiv:2107.12398}}.

\bibitem{LopesCardoso:1998tkj}
G.~Lopes~Cardoso, B.~de~Wit, and T.~Mohaupt, {\it {Corrections to macroscopic
  supersymmetric black hole entropy}},  {\em Phys. Lett. B} {\bf 451} (1999)
  309--316, [\href{http://arxiv.org/abs/hep-th/9812082}{{\tt hep-th/9812082}}].

\bibitem{LopesCardoso:1999fsj}
G.~Lopes~Cardoso, B.~de~Wit, and T.~Mohaupt, {\it {Macroscopic entropy formulae
  and nonholomorphic corrections for supersymmetric black holes}},  {\em Nucl.
  Phys. B} {\bf 567} (2000) 87--110,
  [\href{http://arxiv.org/abs/hep-th/9906094}{{\tt hep-th/9906094}}].

\bibitem{LopesCardoso:2000qm}
G.~Lopes~Cardoso, B.~de~Wit, J.~Kappeli, and T.~Mohaupt, {\it {Stationary BPS
  solutions in N=2 supergravity with R**2 interactions}},  {\em JHEP} {\bf 12}
  (2000) 019, [\href{http://arxiv.org/abs/hep-th/0009234}{{\tt
  hep-th/0009234}}].

\bibitem{Hristov:2022pmo}
K.~Hristov, {\it {The dark (BPS) side of thermodynamics in Minkowski$_{4}$}},
  {\em JHEP} {\bf 09} (2022) 204, [\href{http://arxiv.org/abs/2207.12437}{{\tt
  arXiv:2207.12437}}].

\bibitem{Aharony:2008ug}
O.~Aharony, O.~Bergman, D.~L. Jafferis, and J.~Maldacena, {\it {N=6
  superconformal Chern-Simons-matter theories, M2-branes and their gravity
  duals}},  {\em JHEP} {\bf 10} (2008) 091,
  [\href{http://arxiv.org/abs/0806.1218}{{\tt arXiv:0806.1218}}].

\bibitem{Fuji:2011km}
H.~Fuji, S.~Hirano, and S.~Moriyama, {\it {Summing Up All Genus Free Energy of
  ABJM Matrix Model}},  {\em JHEP} {\bf 08} (2011) 001,
  [\href{http://arxiv.org/abs/1106.4631}{{\tt arXiv:1106.4631}}].

\bibitem{Marino:2011eh}
M.~Marino and P.~Putrov, {\it {ABJM theory as a Fermi gas}},  {\em J. Stat.
  Mech.} {\bf 1203} (2012) P03001, [\href{http://arxiv.org/abs/1110.4066}{{\tt
  arXiv:1110.4066}}].

\bibitem{Nosaka:2015iiw}
T.~Nosaka, {\it {Instanton effects in ABJM theory with general R-charge
  assignments}},  {\em JHEP} {\bf 03} (2016) 059,
  [\href{http://arxiv.org/abs/1512.02862}{{\tt arXiv:1512.02862}}].

\bibitem{Hatsuda:2016uqa}
Y.~Hatsuda, {\it {ABJM on ellipsoid and topological strings}},  {\em JHEP} {\bf
  07} (2016) 026, [\href{http://arxiv.org/abs/1601.02728}{{\tt
  arXiv:1601.02728}}].

\bibitem{Chester:2021gdw}
S.~M. Chester, R.~R. Kalloor, and A.~Sharon, {\it {Squashing, Mass, and
  Holography for 3d Sphere Free Energy}},  {\em JHEP} {\bf 04} (2021) 244,
  [\href{http://arxiv.org/abs/2102.05643}{{\tt arXiv:2102.05643}}].

\bibitem{Bobev:2022jte}
N.~Bobev, J.~Hong, and V.~Reys, {\it {Large N Partition Functions, Holography,
  and Black Holes}},  {\em Phys. Rev. Lett.} {\bf 129} (2022), no.~4 041602,
  [\href{http://arxiv.org/abs/2203.14981}{{\tt arXiv:2203.14981}}].

\bibitem{Bobev:2022wem}
N.~Bobev, S.~Choi, J.~Hong, and V.~Reys, {\it {Large $N$ Superconformal Indices
  for 3d Holographic SCFTs}},  \href{http://arxiv.org/abs/2210.15326}{{\tt
  arXiv:2210.15326}}.

\bibitem{Hosseini:2022vho}
S.~M. Hosseini and A.~Zaffaroni, {\it {The large N limit of topologically
  twisted indices: a direct approach}},  {\em JHEP} {\bf 12} (2022) 025,
  [\href{http://arxiv.org/abs/2209.09274}{{\tt arXiv:2209.09274}}].

\bibitem{Benini:2015eyy}
F.~Benini, K.~Hristov, and A.~Zaffaroni, {\it {Black hole microstates in
  AdS$_{4}$ from supersymmetric localization}},  {\em JHEP} {\bf 05} (2016)
  054, [\href{http://arxiv.org/abs/1511.04085}{{\tt arXiv:1511.04085}}].

\bibitem{Faedo:2021nub}
F.~Faedo and D.~Martelli, {\it {D4-branes wrapped on a spindle}},  {\em JHEP}
  {\bf 02} (2022) 101, [\href{http://arxiv.org/abs/2111.13660}{{\tt
  arXiv:2111.13660}}].

\bibitem{Boido:2022mbe}
A.~Boido, J.~P. Gauntlett, D.~Martelli, and J.~Sparks, {\it {Gravitational
  Blocks, Spindles and GK Geometry}},
  \href{http://arxiv.org/abs/2211.02662}{{\tt arXiv:2211.02662}}.

\bibitem{Robles-Llana:2006vct}
D.~Robles-Llana, F.~Saueressig, and S.~Vandoren, {\it {String loop corrected
  hypermultiplet moduli spaces}},  {\em JHEP} {\bf 03} (2006) 081,
  [\href{http://arxiv.org/abs/hep-th/0602164}{{\tt hep-th/0602164}}].

\bibitem{Antoniadis:2003sw}
I.~Antoniadis, R.~Minasian, S.~Theisen, and P.~Vanhove, {\it {String loop
  corrections to the universal hypermultiplet}},  {\em Class. Quant. Grav.}
  {\bf 20} (2003) 5079--5102, [\href{http://arxiv.org/abs/hep-th/0307268}{{\tt
  hep-th/0307268}}].

\bibitem{Anguelova:2004sj}
L.~Anguelova, M.~Rocek, and S.~Vandoren, {\it {Quantum corrections to the
  universal hypermultiplet and superspace}},  {\em Phys. Rev. D} {\bf 70}
  (2004) 066001, [\href{http://arxiv.org/abs/hep-th/0402132}{{\tt
  hep-th/0402132}}].

\bibitem{Hristov:2010eu}
K.~Hristov, H.~Looyestijn, and S.~Vandoren, {\it {BPS black holes in N=2 D=4
  gauged supergravities}},  {\em JHEP} {\bf 08} (2010) 103,
  [\href{http://arxiv.org/abs/1005.3650}{{\tt arXiv:1005.3650}}].

\bibitem{Hosseini:2017fjo}
S.~M. Hosseini, K.~Hristov, and A.~Passias, {\it {Holographic microstate
  counting for AdS$_{4}$ black holes in massive IIA supergravity}},  {\em JHEP}
  {\bf 10} (2017) 190, [\href{http://arxiv.org/abs/1707.06884}{{\tt
  arXiv:1707.06884}}].

\bibitem{Hosseini:2020vgl}
S.~M. Hosseini, K.~Hristov, Y.~Tachikawa, and A.~Zaffaroni, {\it {Anomalies,
  Black strings and the charged Cardy formula}},  {\em JHEP} {\bf 09} (2020)
  167, [\href{http://arxiv.org/abs/2006.08629}{{\tt arXiv:2006.08629}}].

\bibitem{Ferrero:2020laf}
P.~Ferrero, J.~P. Gauntlett, J.~M. P\'erez Ipi\~na, D.~Martelli, and J.~Sparks,
  {\it {D3-Branes Wrapped on a Spindle}},  {\em Phys. Rev. Lett.} {\bf 126}
  (2021), no.~11 111601, [\href{http://arxiv.org/abs/2011.10579}{{\tt
  arXiv:2011.10579}}].

\bibitem{Ferrero:2020twa}
P.~Ferrero, J.~P. Gauntlett, J.~M.~P. Ipi\~na, D.~Martelli, and J.~Sparks, {\it
  {Accelerating Black Holes and Spinning Spindles}},
  \href{http://arxiv.org/abs/2012.08530}{{\tt arXiv:2012.08530}}.

\bibitem{Hosseini:2021fge}
S.~M. Hosseini, K.~Hristov, and A.~Zaffaroni, {\it {Rotating multi-charge
  spindles and their microstates}},
  \href{http://arxiv.org/abs/2104.11249}{{\tt arXiv:2104.11249}}.

\end{thebibliography}\endgroup

\end{document}